\documentclass[3p]{elsarticle}
\usepackage{epstopdf}
\usepackage{subfigure,subcaption,graphicx,float}
\usepackage{amsmath, amsfonts, amssymb}
\usepackage{amsthm}
\usepackage{yhmath}
\usepackage{epsf}
\usepackage{amsfonts}
\usepackage{stmaryrd}
\usepackage{amssymb}
\usepackage{leftidx}
\usepackage{xcolor}
\usepackage{mathtools}
\usepackage{placeins}
\usepackage{booktabs}
\usepackage{enumitem}
\usepackage{caption}
\usepackage{tabu}
\usepackage{multirow}
\usepackage{stackengine}
\usepackage[utf8]{inputenc}
\usepackage[english]{babel}
\usepackage{bm}
\usepackage{array}
\usepackage{multirow}

\def\bfb{{\bf b}}

\def\bfu{{\bf u}}

\def\bfE{{\bf E}}

\def\bfI{{\bf I}}

\def\bfX{{\bf X}}

\def\eps{\varepsilon}

\def\e0{\varepsilon_0}

\def\s0{\sigma_0}

\def\sts{\sigma_{\texttt{ts}}}
\def\scs{\sigma_{\texttt{cs}}}
\def\shs{\sigma_{\texttt{hs}}}
\def\de{\delta^\varepsilon}
\def\ce{c_{\texttt{e}}}

\DeclareMathAlphabet{\mathsfit}{T1}{\sfdefault}{\mddefault}{\sldefault}
\SetMathAlphabet{\mathsfit}{bold}{T1}{\sfdefault}{\bfdefault}{\sldefault}

\theoremstyle{plain}
\newtheorem{theorem}{Theorem}

\newtheorem{remark}[theorem]{Remark} 

\long\def\symbolfootnote[#1]#2{\begingroup%
\def\thefootnote{\fnsymbol{footnote}}\footnote[#1]{#2}\endgroup}

\begin{document}
\begin{frontmatter}



\title{Dynamic phase-field model for brittle fracture in grounded glaciers \vspace{0.1cm}}

\vspace{-0.1cm}

\author[GT]{Aarosh Dahal}
\ead{adahal8@gatech.edu}

\author[GT]{Umar Khayaz}
\ead{ukhayaz3@gatech.edu}

\author[Vanderbilt]{Ravindra Duddu}
\ead{ravindra.duddu@vanderbilt.edu}

\author[GT]{Aditya Kumar\corref{cor1}}
\ead{aditya.kumar@ce.gatech.edu}

\address[GT]{School of Civil and Environmental Engineering, Georgia Institute of Technology, Atlanta, GA 30332, USA \vspace{0.05cm}}

\address[Vanderbilt]{Department of Civil and Environmental Engineering, Vanderbilt University,  Nashville, TN 37235, USA \vspace{0.05cm}}

\cortext[cor1]{Corresponding author}

\begin{abstract}

\vspace{-0.1cm}

Fracture and calving of glaciers are key contributors to ice-mass loss and sea-level rise, yet predictive modeling remains challenging. Fracture in grounded glaciers is driven by gravitational forces and is typically studied within the framework of quasi-static linear elastic fracture mechanics. In this work, we show that purely quasistatic brittle fracture simulations within the phase field fracture framework under fixed self-weight can become strongly overdriven after crevasse initiation, producing unphysical thickening of the diffusive crack band and diffuse damage patterns. This pathology arises because gravity drives a growing region ahead of the crack tip beyond the strength surface. To resolve this, we show that the post-nucleation propagation is fundamentally a dynamic instability rather than a quasistatic process and propose the use of dynamic formulations of fracture. We demonstrate that accounting for inertia results in sharp, localized cracks that propagate through the ice thickness. As a second objective, this paper introduces a new dynamic formulation of the phase-field fracture model of Kumar et al. (J. Mech. Phys. Solids 112:523–551, 2018) in which elastic, inertial, and gravitational contributions are degraded consistently in fractured regions.

\endkeyword

\end{abstract}

\end{frontmatter}

\section{Introduction}

Fracture and calving of glaciers, resulting in mass loss, are among the primary contributors to global sea-level rise \cite{meier2007glaciers}. 
Glaciers can be broadly distinguished by whether their downstream end is floating or grounded. In floating extensions, such as ice shelves and ice tongues, the ice extends over the ocean and is supported by buoyancy. In grounded glaciers, the ice remains in contact with the underlying bed, either on land or below sea level.
While fracture of floating ice shelves has received considerable attention, grounded glaciers present a distinct and less well-understood regime in which the full weight of the ice is supported by the bed, producing stress fields that are strongly influenced by gravity. 
The flow and stability of grounded glaciers are strongly influenced by processes near the terminus, where crevassing can trigger calving and rapid retreat.  
Understanding the mechanisms that drive full-depth fractures is therefore essential for predicting the future evolution of glaciers and their contribution to sea-level change \cite{scambos2000link, bassis2013diverse, benn2017glacier}.

Observations indicate that fracture nucleation often initiates at surface or basal crevasses formed due to tensile stretching, bending near the terminus, buoyancy-related flexure for marine-terminating settings, or hydrological processes \cite{vaughan1993relating, van1998fracture, kingslake2017widespread}. Once initiated, these cracks can propagate rapidly through the ice thickness. 
The long-term deformation of glaciers is dominated by viscous or viscoplastic creep, crack nucleation and gradual growth may occur over days to years as stresses evolve. However, the final stage of propagation, particularly the through-thickness fracture that precipitates calving, can occur on timescales of seconds to minutes, far shorter than those of viscous flow. On these short timescales, ice behaves as a brittle elastic solid, and the propagation event can be treated within the framework of elastic brittle fracture mechanics \cite{pralong2005dynamic}.

Several modeling approaches have been proposed to study glacier fracture. 
Early studies on glacier fracture relied on Linear Elastic Fracture Mechanics (LEFM) frameworks \cite{van1998fracture, weiss2004subcritical}. While such methods provide useful insights into critical crack lengths and propagation conditions, they require additional empirical criteria to predict crack geometry and cannot predict crack nucleation. More recently, the phase-field approach to brittle fracture has emerged as a powerful alternative. The classical phase field model \cite{bourdin2000numerical} is constructed as a regularization of the variational theory of brittle fracture of Francfort and Marigo \cite{Francfort98}. This theory argues that, under quasi-static loading, both when and where a crack grows can be obtained from Griffith’s fracture postulate in its general form as an energy cost-benefit principle. More precisely, it states that the displacement field $\mathbf{u}(\mathbf{X},t)$ and the crack set $\Gamma(t)$, under quasi-static displacement-controlled loading, are obtained by globally minimizing the total energy, defined as the sum of elastic and fracture contributions:
\begin{equation}
 \mathcal{E}(\mathbf{u}, \Gamma) := \int_{\Omega_0 \setminus \Gamma} W(\mathbf{u}) \, \mathrm{d}\mathbf{X} 
+ G_c \, \mathcal{A}(\Gamma), 
\label{Variational} 
\end{equation}
where $\mathcal{A}(\Gamma)$ stands for the surface measure (2–dimensional Hausdorff measure) of the unknown crack and $W(\mathbf{u})$ stands for the hyperelastic energy function.
This formulation, in its regularized phase-field form, provides a mathematically rigorous, physically consistent, and numerically robust method for simulating crack growth in two or three dimensions. Although the classical model does not explicitly include material strength, its numerical implementation requires choosing a finite regularization length, which in turn induces a regularization-length-dependent effective strength.

In the context of ice mechanics, the classical phase-field model has been applied by Sun {et al.} \cite{sun2021poro} to simulate crevasse growth and calving. They observed the formation of diffuse damage patterns that bear little resemblance to the sharp, localized fractures observed in nature. Their parametric study showed that localized cracks are obtained only when a very high material strength is adopted. Since this requires a very small regularization length, making the computations prohibitively expensive, they proposed a computational modification based on introducing an energy threshold into the model, following the work of Miehe {et al.} \cite{miehe2015}. Specifically, the energy functional (\ref{Variational}) is modified as
\begin{equation}
 \mathcal{E}^{*}(\mathbf{u}, \Gamma) := \int_{\Omega_0 \setminus \Gamma} \left(W(\mathbf{u}) - W^{*} \right) \, \mathrm{d}\mathbf{X} 
+ G_c \, \mathcal{A}(\Gamma), 
\label{Variational-energylimiter} 
\end{equation}
where $ W^{*}$ is a pre-set constant energy threshold. Later, to incorporate the long-term creep behavior of glacial ice, Clayton et al. \cite{clayton2022stress} proposed a non-variational formulation for modeling fracture in a viscoplastic material. While this formulation introduced several changes to the classical elastic brittle fracture framework used by Sun {et al.}, it retained a key ingredient: the use of an energy threshold function to induce damage localization. Other work on modeling viscoelastic fracture includes that of Sonderhaus et al. \cite{sondershaus2023phase}. However, the underlying principles governing failure in viscoelastic fluids under general loading conditions are not yet well established, which makes regularized phase-field models for such materials somewhat \textit{ad hoc}.

Recent work by Khayaz {et al.} \cite{khayaz2025comparison} has shown that the modifications introduced by Miehe et al. \cite{miehe2015} to the classical variational model disturb the Griffith energetic competition. Moreover, glacier ice is a brittle material with finite tensile and compressive strength at large strain rates. The absence of a proper strength-based nucleation criterion therefore limits the accuracy of phase-field predictions under homogeneously distributed stress states, such as those generated by gravity loading. Furthermore, for nearly incompressible materials, both classical phase-field formulations and Miehe-type alternatives are known to become ill-behaved under hydrostatic tensile states \cite{KFLP18, KBFLP20}.

To overcome these limitations and incorporate the strength of ice directly into the model, we employ the phase-field approach developed by Kumar {et al.} \cite{KFLP18, kumar2020revisiting}. In a nutshell, the theory of Kumar {et al.} corresponds to a generalization of the classical variational phase field models of brittle fracture by accounting for the strength of the material at large. The strength of an elastic brittle material is the set of all critical stresses $\boldsymbol{\sigma}$ at which the material fractures when it is subjected to a state of monotonically increasing, spatially uniform, but otherwise arbitrary stress. Such a set of critical stresses defines a surface in stress space
\begin{equation*}
\mathcal{F}(\boldsymbol{\sigma})=0,
\end{equation*}
called the strength surface of the material. The generalized model dictates that violation of the strength surface is a necessary condition for crack nucleation. Under uniform stresses, this condition is also sufficient. In the presence of stress concentrators, however, the Griffith variational problem (\ref{Variational}) must also be satisfied. This model has been widely validated across a range of problems \cite{KLP21, KRLP22, KLDLP23, LK24, KKLP24, kamarei2026nine, WK2025} and has been shown to accurately predict arbitrary crack nucleation and propagation in both soft and hard elastic brittle materials under quasi-static loading.

In this work, we first revisit the quasi-static analysis of gravity-driven fracture in grounded glaciers using the phase-field model of Kumar {et al.}
We adopt realistic material strength values for glacial ice rather than choosing \textit{ad hoc} values.
We show that for grounded glaciers, where the gravitational body force continuously overdrives the stress state well beyond the strength surface once a crack has initiated, quasistatic formulation leads to a progressive, unphysical thickening of the phase-field crack band and, ultimately, diffuse damage patterns observed in previous work \cite{sun2021poro}.
This pathology is not specific to any one phase-field variant; we demonstrate that it arises both in the complete nucleation model and in classical variational models with volumetric–deviatoric energy splits.

This leads to the central insight of this work: a dynamic formulation is essential for correctly capturing gravity-driven glacier fracture. Because the gravitational pre-stress is large relative to the material strength and the resulting calving events are rapid, with reported timescales on the order of 10--100 ms \cite{graff2025calving}, post-nucleation propagation is fundamentally dynamic. This is an unusual dynamic fracture problem because dynamic fracture is typically initiated by external loads applied over short durations. Here, instead, dynamic fracture is driven by the introduction of a stress concentrator into a material that was previously stable under substantial pre-stress. A similar phenomenon can be observed in the explosive fragmentation of a Prince Rupert's drop \cite{kooij2021prince}.

The dynamic fracture model used here is based on the work of Liu {et al.} \cite{liu2024dynamic}, who extended the quasi-static formulation of Kumar {et al.} to the dynamic setting. We introduce one important modification concerning the treatment of mass density in the regularized crack region.

Extensions of quasi-static classical variational phase field models and cohesive phase field models for brittle fracture to the dynamic setting have been extensively studied \cite{Bourdin2011_TimeDiscreteDynamic, Borden2012_PhaseFieldDynamicBrittle, Hofacker2012_ContinuumPhaseField, Bleyer2017_DynamicCrackVariational, Nguyen2018_PhaseFieldCohesive, Geelen2019_PhaseField, Mandal2020_EvaluationVariational}. Several questions remain open in dynamic phase-field fracture, including the dependence of critical energy release rate and strength on crack velocity, and the ability of the models to predict specific features of dynamic crack growth, such as limiting crack speed and crack branching. However, one feature common to previous dynamic fracture extensions is that the kinetic energy, or equivalently the mass density, is left undegraded in the regularized crack region.

The usual motivation for this constitutive choice is mass conservation \cite{Borden2012_PhaseFieldDynamicBrittle, marigo2016dynamics}. Although phase-field fracture models were developed as regularizations of variational fracture, they have often been interpreted as gradient damage models. In that interpretation, the phase field is a macroscopic damage variable. The regularization length is then interpreted as a material length scale and can no longer be taken as an arbitrarily small parameter.  Within this gradient-damage interpretation, leaving the density undegraded is natural.

However, undegraded density allows the regularized crack region to carry momentum and inertia even as its stiffness and, consequently, elastic wave speed vanish. As a result, kinetic energy persists inside the damage band and can drive artificial crack-band widening, a phenomenon reported in several dynamic phase-field studies \cite{Bleyer2017_DynamicCrackVariational, Mandal2020_EvaluationVariational}, with some even prescribing physical meaning to it \cite{Mandal2020_EvaluationVariational}.
We argue that this widening is a numerical artifact that can produce spurious effects, including artificially reduced crack speeds.

In contrast, when the phase-field model is viewed as a regularized \emph{fracture} model, degrading density introduces only a regularization error in mass conservation, analogous to the error introduced by smearing a sharp crack over a finite length scale. This error vanishes as the regularization length decreases. The phase-field approach of Kumar et al. adopts this regularized-fracture interpretation: the regularization length is a numerical parameter that can be made arbitrarily small while the physical strength surface is retained. We therefore compare degraded and undegraded density formulations within this approach. Benchmark simulations, including the Kalthoff--Winkler test and a dynamic branching test, are used to demonstrate the spurious effects of undegraded density, and a one-dimensional bar study is used to identify the condition required for the regularized crack to behave as an open crack.

The organization of the paper is as follows. Section \ref{Sec: Problem} introduces the problem of crevassing and fracture in grounded glaciers. The quasi-static analysis is presented in Section \ref{Sec:quasistatic}. The dynamic fracture tests to compare the undegraded and degraded density formulations are presented in Section \ref{Sec: Dynamic}, followed by the dynamic analysis of crevassing in glaciers. We conclude this work by recording a number of final comments in Section \ref{Sec: Final Comments}.

\section{The problem}\label{Sec: Problem}

\begin{figure}[h!]
	\centering
	\includegraphics[width=6.3in]{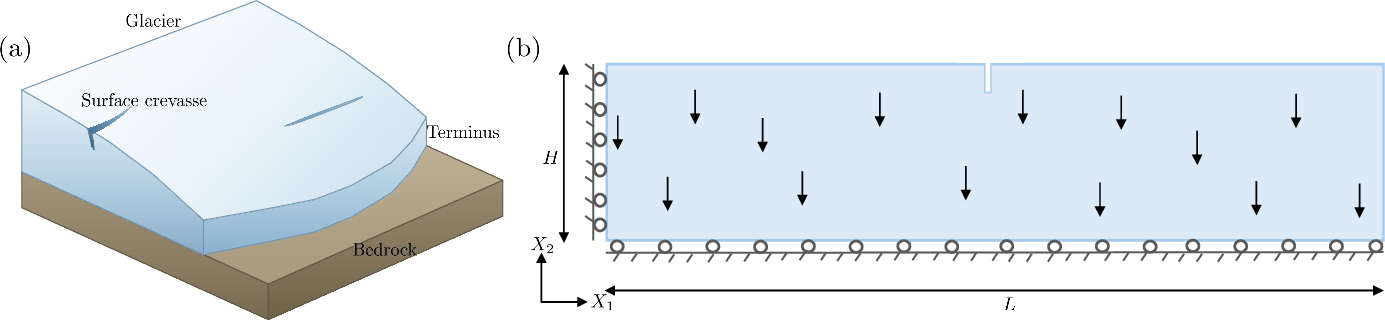}
	\caption{ (a) Illustration of a land-terminating grounded glacier with surface crevasses, and (b) a schamatic showing the boundary conditions of a grounded glacier of length $L=300$ m and height $H=75$ m containing a single surface crevasse of depth 1 m}\label{Fig1}
\end{figure}

We consider a land-terminating grounded glacier whose right terminus is traction-free and whose left boundary remains connected to the ice sheet, as illustrated in Fig.~\ref{Fig1}. The glacier forms over long timescales through the gradual accumulation and compaction of snow and ice, giving the glacial ice unique properties. Glacial (freshwater) ice exhibits complex rheology that is most accurately described as viscoplastic over long time scales \cite{montagnat2004viscoplastic}. In contrast, fracture and calving events occur in short timescales. At high strain rates, laboratory observations indicate that glacier ice responds in a predominantly brittle manner \cite{deng2020experimental}. Motivated by this separation of time scales, we idealize the material as a brittle elastic solid during fracture events.

Stresses within the grounded glacier arise primarily from gravitational body forces associated with its substantial self-weight, although hydrological forcing near the terminus may generate additional localized stresses.
Fracture processes are usually triggered by the formation of crevasses. Such crevasses usually form due to surface-water melting \cite{colgan2016glacier}, flexural fatigue \cite{ren2014effects}, tidal bending \cite{hulbe2016tidal}, etc. However, the exact mechanism is not the focus of this work, but rather its consequences. We assume that once a sufficiently deep crevasse exists, the subsequent fracture and calving dynamics are governed by gravity–driven stresses.

\subsection{Initial configuration, and kinematics}

Since the thickness of a glacier is large relative to its length and width, we can use two-dimensional plane-strain analysis. 
Consider a two-dimensional domain representing the glacier occupying an open bounded domain $\mathrm{\Omega}\subset \mathbb{R}^3$, with boundary $\partial\mathrm{\Omega}$ in its undeformed and stress-free configuration at time $t=0$. 
The geometry and the boundary conditions are shown in Fig.~\ref{Fig1}(b). Free-slip boundary conditions, indicated by rollers in the figure, are applied along the left and bottom edges, referred to as $\partial\mathrm{\Omega}_\mathcal{D}$, and zero traction is applied on the complementary part $\partial\mathrm{\Omega}_\mathcal{N}=\partial\mathrm{\Omega}\setminus \partial\mathrm{\Omega}_\mathcal{D}$.

Under gravitational body forces,  the structure experiences a displacement field $\bfu(\bfX,t)$. We write the infinitesimal strain tensor as
\begin{equation*}
\bfE(\bfu)=\dfrac{1}{2}(\nabla\bfu+\nabla\bfu^T).
\end{equation*}
The elasticity for glacial ice at high rates can be characterized by the isotropic linear elastic stored-energy function
\begin{equation}
	W(\bfE(\bfu)) =\mu \, {\rm tr}\,\bfE^2+\dfrac{\lambda}{2}({\rm tr}\,\bfE)^2,\label{W-mu}
\end{equation}
where $\mu>0$ and $\lambda>-2/3\mu$ are the Lam\'e constants. Recall the basic relations $\mu=E/(2(1+\nu))$ and $\lambda=E\nu/((1+\nu)(1-2\nu))$, where $E$ is the Young's modulus and $\nu$ is the Poisson's ratio. The stress-strain relation is given by
\begin{equation*}
	\boldsymbol{\sigma}(\bfX,t)=\dfrac{E}{1+\nu}\bfE+\dfrac{E\,\nu}{(1+\nu)(1-2\nu)}({\rm tr}\,\bfE)\bfI.
\end{equation*}
Two material properties need to be known to characterize the brittle fracture behavior: (\emph{i}) critical energy release rate and (\emph{ii}) strength.
The critical energy release rate (or fracture toughness), denoted as $G_c$, controls the nucleation from a large pre-existing crack, $\Gamma$, through the Griffith criterion
\begin{equation}
-\dfrac{\partial{\mathcal{W}}}{\partial \Gamma}=G_c \label{Griffith},
\end{equation}
which describes a competition between bulk elastic energy and surface fracture energy. As discussed in the Introduction, the strength of the material controls the nucleation in large specimens of homogeneous brittle materials subjected to a uniform state of stress $\boldsymbol{\sigma}$. In greater than one dimension, the set of critical stresses defines a strength surface in the stress space represented as
\begin{equation}
\mathcal{F}(\boldsymbol{\sigma})=0,\label{SSurf-0}
\end{equation}
A crack will form in an indeterminate location in the homogeneous specimen subjected to arbitrary uniform loading once the stress hits the strength surface in any direction. Under non-uniform stress states, experimental observations have shown that the violation of the strength surface is a necessary but not a sufficient condition for crack nucleation. In that case, an \emph{interpolation} of the strength criterion (\ref{SSurf-0}) and toughness criterion (\ref{Griffith}) defines the fracture nucleation. The question of how the crack, once nucleated, chooses its path is more debatable and will be discussed later.

A popular choice for the strength surface of the material that we will invoke in this work is the Drucker-Prager strength surface
\begin{equation}
	\mathcal{F}(\boldsymbol{\sigma})=\sqrt{J_2}+\gamma_1 I_1+\gamma_0=0\qquad {\rm with}\qquad \left\{\hspace{-0.1cm}\begin{array}{l}\gamma_0=-\dfrac{2\sigma_{\texttt{cs}}\sigma_{\texttt{ts}}}
		{\sqrt{3}\left(\sigma_{\texttt{cs}}+\sigma_{\texttt{ts}}\right)}\vspace{0.2cm}\\
		\gamma_1=\dfrac{\sigma_{\texttt{cs}}-\sigma_{\texttt{ts}}}
		{\sqrt{3}\left(\sigma_{\texttt{cs}}+\sigma_{\texttt{ts}}\right)}\end{array}\right. ,\label{DP}
\end{equation}
where
\begin{equation}
	I_1={\rm tr}\,\boldsymbol{\sigma}\qquad {\rm and}\qquad  J_2=\dfrac{1}{2}{\rm tr}\,\boldsymbol{\sigma}^2_{D}\qquad {\rm with}\quad \boldsymbol{\sigma}_{D}=\boldsymbol{\sigma}-\dfrac{1}{3}({\rm tr}\,\boldsymbol{\sigma})\bfI\label{T-invariants}
\end{equation}
stand for two of the standard invariants of the stress tensor $\boldsymbol{\sigma}$, while the constants $\sigma_{\texttt{ts}}>0$ and $\sigma_{\texttt{cs}}>0$ denote the uniaxial tensile and compressive strengths of the material. We emphasize that this form of the strength surface is a constitutive modeling choice, as experimental data are typically available at only a limited number of stress states. Alternative representations, such as the Mohr–Coulomb criterion, may be adopted without additional difficulty when sufficient data are available to calibrate them reliably.

Accepting this description of material behavior, the central task lies in how we write down the balance equations that capture both the strength criterion (\ref{SSurf-0}) and the toughness criterion (\ref{Griffith}), along with their interpolation, thereby completely defining fracture nucleation. This formulation is presented in Section \ref{Sec:quasistatic}. Before doing so, we summarize the material properties of glacial ice adopted from the literature.

\subsection{Material parameters for glacial ice}

We list the material properties of the glacial ice in Table~\ref{tab:material-ice}.
The values for the Young's modulus, $E$, and Poisson's ratio, $\nu$, are taken from Duddu and Waisman \cite{duddu2012temperature}.  It is to be noted that the properties of ice are heavily rate-dependent \cite{hawkes1972deformation}. For instance, Murat {et al}. \cite{murat1982some} reports a range of Poisson's ratio of sea-ice from close to 0.5 at low strain rates to around 0.35 at high strain-rates. We adopt the high strain rate value for this work, although note that the phase field formulation presented in the following sections can handle near incompressibility. The densities of glacier ice and seawater are adopted from Jimenez and Duddu \cite{jimenez2018evaluation}. The critical Mode I stress intensity factor, $K_c$, for glacier ice is taken from the tests of Liu et al. \cite{liu1979fracture} which allows for the calculation of $G_c$. For strength values, we rely on the work of Petrovic (2003) \cite{petrovic2003review}, who reports ice tensile strength, $\sts$, and compressive strength, $\scs$, in the range 0.7 MPa to 3.1 MPa and 5 MPa to 25 MPa respectively. For computational convenience in the simulations presented below, we adopt values near the lower end of the reported ranges for both parameters; however, the results and the conclusions are not meaningfully affected by this choice in our trials.  

\begin{table}[h!]
\centering
\caption{Material constants for glacial ice used throughout this work.\label{tab:material-ice}}
\begin{tabular}{lcc}
\hline
\textbf{Constant} & \textbf{Symbol} & \textbf{Value} \\
\hline
Mass density of ice & $\rho$ & $917~\mathrm{kg/m^3}$ \\
Young's modulus & $E$ & $9.5 \times 10^9~\mathrm{N/m^2}$ \\
Poisson's ratio & $\nu$ & $0.35$ \\
Mode I stress intensity factor & $K_c$ & $0.1 \times 10^6~\mathrm{Nm^{-3/2}}$ \\
Critical energy release rate & $G_c$ & $0.924~\mathrm{J/m^2}$ \\
Uniaxial tensile strength & $\sigma_{\texttt{ts}}$ & $0.7 \times 10^6~\mathrm{N/m^2}$ \\
Uniaxial compressive strength & $\sigma_{\texttt{cs}}$ & $5.0 \times 10^6~\mathrm{N/m^2}$ \\
\hline
\end{tabular}
\end{table}

\section{Analysis with a quasi-static fracture model}\label{Sec:quasistatic}

\subsection{The complete phase-field theory of fracture nucleation and propagation}

\noindent In the phase field approach, a scalar field
\begin{equation*}
v=v(\bfX,t)
\end{equation*}
is introduced to regularize the crack surface, which takes values in the range [0, 1] over a phase boundary of infinitesimal width $\varepsilon$. Precisely, $v = 1$ identifies regions of the sound material, whereas $v < 1$ identifies regions of the material that have been fractured.

In the Kumar et al. phase-field model \cite{KFLP18, KBFLP20}, the displacement field $\bfu_k(\bfX)=\bfu(\bfX,t_k)$ and phase-field $v_k(\bfX)=v(\bfX,t_k)$ at any material point $\bf X\in\overline{\mathrm{\Omega}}$ and discrete time $t_k\in\{0=t_0,t_1,...,t_m,$ $t_{m+1},...,$ $t_M=T\}$ are determined by the system of coupled partial differential equations (PDEs)
\begin{equation}
\left\{\begin{array}{ll}
\hspace{-0.15cm} {\rm Div}\left[(v_{k}^2 +\eta)\dfrac{\partial W}{\partial \bfE}(\bfE(\bfu_{k}))\right]+(v_{k}^2 + \eta)\bfb={\bf0} ,
\vspace{0.1cm}\\
\hspace{-0.15cm}
\dfrac{3}{4} \varepsilon \, \de \,  G_c \triangle v_{k}=2 v_{k} W(\bfE(\bfu_{k}))+c_\texttt{e}(\bfX,t_{k})- \dfrac{3}{8}  \dfrac{\de \, G_c}{\varepsilon}+\dfrac{\zeta}{2}\,p\!\left(v_{k-1},v_k\right),
\end{array}\right. \label{phase-field-equations}
\end{equation}
where $p\!\left(v_{k-1},v_k\right) = \mathcal{H}(v_a-v_k)
\left(|v_{k-1}-v_k|-(v_{k-1}-v_k)\right) +|1-v_k|-(1-v_k) -|v_k|+v_k$ is the penalty function that ensures the phase field remains in the physically admissible range $0 \leq v \leq 1$ and that fracture is irreversible for $v<v_a$. The value of $v_a$ is set to 0.05 because this allows the crack to fully develop and yield the most physical results \cite{kumar2020revisiting, dolbow2025uniform}. The parameter $\zeta$ is typically set to $10^{3} {\de \, G_c}/{\varepsilon}$. The constant $\eta$, which contributes to a small residual stiffness in cracked regions, is set to $10^{-7}$ but can be set to 0 as well in the small deformation limit.

\noindent The constitutive prescription for the driving force $c_\texttt{e}(\bfX,t)$ in (\ref{phase-field-equations})$_2$ as provided by Kumar et al. \cite{KRLP22} is 
\begin{align}
c_{\texttt{e}}(\bfX,t)=\beta_2^\varepsilon\sqrt{\mathcal{J}_2}+\beta_1^\varepsilon \mathcal{I}_1-
v (1-sgn(\mathcal{I}_1)) \, W(\bfE(\bfu)),\label{cehat-2022}
\end{align}
with
\begin{equation}
\left\{\begin{array}{l}
\beta^\varepsilon_1=\dfrac{1}{\shs}\delta^\varepsilon\dfrac{G_c}{8\varepsilon}-\dfrac{2\mathcal{W}_{\texttt{hs}}}{3\shs}\vspace{0.2cm}\\
\beta^\varepsilon_2=\dfrac{\sqrt{3}(3\shs-\sts)}{\shs\sts}\delta^\varepsilon\dfrac{G_c}{8\varepsilon}+
\dfrac{2\mathcal{W}_{\texttt{hs}}}{\sqrt{3}\shs}-\dfrac{2\sqrt{3}\mathcal{W}_{\texttt{ts}}}{\sts}\end{array}\right. , \label{betas}
\end{equation}
and
$$\mathcal{W}_{\texttt{ts}}= \dfrac{\sts^2}{2 E}, \quad \mathcal{W}_{\texttt{hs}}= \dfrac{\shs^2}{2 \kappa}, \quad \shs=\dfrac{2 \sts  \scs} { 3 (\scs - \sts)}, \quad \kappa= \dfrac{E}{3 (1 - 2 \nu)}.$$
$\mathcal{I}_1$ and $\mathcal{J}_2$ stand for the invariants (\ref{T-invariants}) of the degraded Cauchy stress
\begin{equation*}
\boldsymbol{\sigma}(\bfX,t)=v^2\dfrac{\partial W}{\partial \bfE}(\bfE(\bfu))
\end{equation*}
and, hence, read as
\begin{equation*}
\mathcal{I}_1=(3\lambda+2\mu) v^2 {\rm tr}\,\bfE(\bfu)\quad {\rm and}\quad \mathcal{J}_2=2\mu^2 v^4 {\rm tr}\,\bfE^2_D(\bfu)
\end{equation*}
with $\bfE_D(\bfu)=\bfE(\bfu)-1/3\left({\rm tr}\,\bfE(\bfu)\right)\bfI$ in terms of the displacement field $\bfu$ and phase-field $v$. The formulas for the two constants $\beta^\varepsilon_1$ and $\beta^\varepsilon_2$ are obtained by fitting the strength surface exactly at two points---in this case, the uniaxial tensile strength and hydrostatic tensile strength.

The value of the effective critical energy release rate can be corrected to match the experimental value, $G_c$, through the parameter $\de$. An approximate analytical formula for $\de$ was recently provided in Kamarei {et al.} \cite{KKLP24}.
\begin{equation}
\delta^\varepsilon=\left(1+\dfrac{3}{8}\dfrac{h}{\varepsilon}\right)^{-2}\left(\dfrac{\sts+(1+2\sqrt{3})\,\shs}{(8+3\sqrt{3})\,\shs}\right)\dfrac{3  G_c}{16 \mathcal{W}_{\texttt{ts}} \varepsilon}+\left(1+\dfrac{3}{8}\dfrac{h}{\varepsilon}\right)^{-1}\dfrac{2}{5}.
\label{delta-eps-final-h}
\end{equation}
where $h$ is the mesh size.  This formula is applicable for any brittle material, whether described by linear elastic or nonlinear elastic theory, for which the strength surface is well captured by the Drucker-Prager strength surface (\ref{DP}). We note again that the theory allows for arbitrary strength surface, not only the Drucker-Prager one, as demonstrated recently by Chockalingam et al. \cite{chockalingam2025MCHB}.

We refer the readers to \cite{KFLP18, KBFLP20} for a full description of the model and the details of numerical implementation. A number of open-source implementations of this model are available \footnote{https://sites.gatech.edu/adityakumar/software-and-presentations/}\textsuperscript{,}\footnote{https://github.com/hugary1995/raccoon.}. 

\begin{remark}{\rm A new feature of the present model is the degradation of the gravitational body force $\bfb = - \rho \textbf{g}$ with the phase field in the balance of linear momentum (\ref{phase-field-equations})$_1$. This choice is motivated by recent results from Nguyen et al. \cite{duddu2025bodyforce}, which showed that the presence of undegraded gravitational force in cracked regions can lead to excessive deformations and numerical convergence issues. We discuss this argument more in Appendix A. We note that this choice is equivalently interpreted as introducing a phase–field–dependent effective density. We will discuss more about the consequences of it in Section \ref{Sec: Dynamic} on dynamic analysis.}
\end{remark}

\begin{remark}{\rm 
In equations (\ref{phase-field-equations}), the phase field can only evolve from its initial value of $v=1$ when
\begin{equation}
\hat{\mathcal{F}}(\boldsymbol{\sigma})=
\dfrac{J_2}{\mu}+\dfrac{I_1^2}{9 \kappa} + \ce - \dfrac{3 \de G_c}{8 \eps}=0 \label{strength-surface-ce}
\end{equation}
obtained by setting $v=1$ in the right-hand side of the evolution equation for phase-field (\ref{phase-field-equations})$_2$.  $\hat{\mathcal{F}}$ can be regarded as a phase-field approximation of the strength surface ${\mathcal{F}}(\boldsymbol{\sigma})$ (\ref{SSurf-0}) and it signals the onset of fracture nucleation.  
The phase-field strength surface reduces to the exact strength surface for $\eps \searrow 0$ \cite{kumar2020revisiting}.  For any material point where
$\hat{\mathcal{F}}(\boldsymbol{\sigma}) < 0$, the stress state lies within the strength surface and the material remains intact. When $\hat{\mathcal{F}}(\boldsymbol{\sigma}) \geq 0$, the local stress state has reached or exceeded the strength surface, and the phase-field is driven to evolve from its initial value. From (\ref{strength-surface-ce}), it is also easy to observe how the introduction of an elastic energy threshold (\ref{Variational-energylimiter}) can artificially increase strength in previous work \cite{sun2021poro}.

Note again that satisfaction of
$\hat{\mathcal{F}}(\boldsymbol{\sigma}) = 0$ at a material point $\mathbf{X}$ is a necessary but not sufficient condition for fracture under non-uniform stress fields. Nevertheless, $\hat{\mathcal{F}}(\boldsymbol{\sigma}) \geq 0$ remains a useful diagnostic for the regions where the phase field is evolving 
in subsequent sections, where we evaluate the stress state ahead of the crack tip to understand the overdriven behavior observed under quasi-static gravitational loading.
}
\end{remark}

\begin{remark}{\rm 
Recent work \cite{LK24, lopez2025Whenandwhere, WK2025} has demonstrated that the phase field equations (\ref{phase-field-equations}) essentially describe a strength-constrained energy minimization and that it accurately predicts the crack path under mixed-mode fracture, including for mode I+III loading where echelon cracking is often observed \cite{WK2025}. Therefore, it resolves the question of the crack path that was left unresolved from Griffith (\ref{Griffith}).
}
\end{remark}

\subsection{Simulations of the glacier}

\begin{figure}[h!]
	\centering
	\includegraphics[width=5.5in]{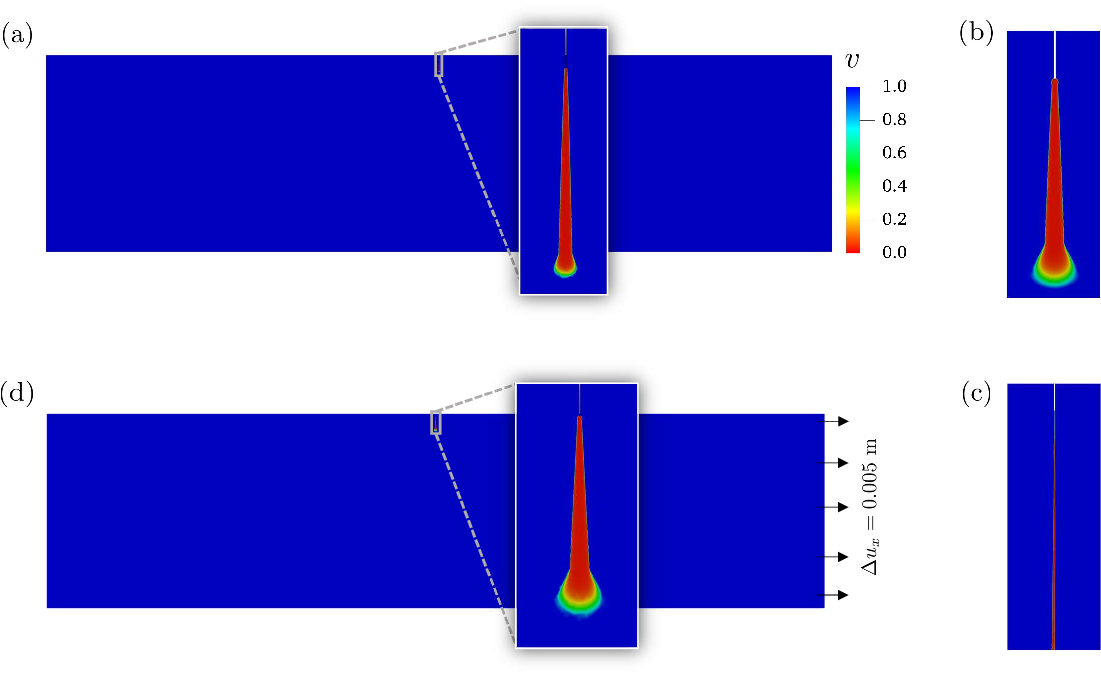}
	\caption{Contour plots of the phase field from quasi-static simulations of a glacier 100 m long and 75 m tall with a surface crevasse 1 m deep and fracturing under self-weight. Results are shown for (a) the phase-field model of Kumar et al. \eqref{phase-field-equations} and (b) the classical variational model with a volumetric-deviatoric energy split \cite{AmorMarigoMaurini2009}. Part (c) shows the result when the body force is artificially reduced by a factor of two. Part (d) shows the result when fracture is driven by a large externally applied load on the right boundary rather than by self-weight.}\label{Fig2}
\end{figure}

Fig.~\ref{Fig2}(a) shows the results of simulating the grounded glacier problem introduced in Section \ref{Sec: Problem} using the quasi-static phase-field fracture model, as described by the governing equations (\ref{phase-field-equations}). The dimensions of the glacier are $L=300$ m and $H=100$ m, and it contains an initial crevasse of depth 1 m. Realistic strength values are adopted as mentioned in Section 2.2. The phase field under the pre-notch start to evolve under the action of gravitational load. Note that there are no other time-dependent loads in this problem.

The results show a peculiar behavior. The phase field starts to progressively thicken as the damaged region migrates downward. In the AT1 regularization, a sharp crack is represented by a phase-field profile with compact support spanning a width of 4$\varepsilon$. It is obvious from the results in Fig.~\ref{Fig2} that the width of the diffused damage zone is constantly increasing and is not representative of a crack. We later plot the crack profile in Fig.~\ref{Fig7} to demonstrate that it is much wider than 4$\varepsilon$.

This behavior of the phase-field method is not a product of the particular phase-field model being used; similar behavior is also seen for the classical variational model with the volumetric-deviatoric energy split \cite{AmorMarigoMaurini2009}, as shown in Fig.~\ref{Fig2}(b), where we adjusted the value of parameter $\varepsilon$ so that the same tensile strength of 0.7 MPa is embedded in the model. Also, it is worth noting that the thickening is less severe when the gravitational body force is artificially reduced, as shown in Fig.~\ref{Fig2}(d), indicating that overdriven stress conditions drive diffuse damage. 

A similar behavior is also observed in a test in which we remove the gravitational body force and instead apply a stretch on the boundary in one time step that is well beyond the critical displacement at which the pre-crack will start to propagate, as shown in Fig.~\ref{Fig2}(c).

As we noted in the Introduction, this problem has been recognized in previous work \cite{sun2021poro}. However, the underlying cause of this behavior was not identified. Instead, a computational ``fix'' of introducing a energy threshold (\ref{Variational-energylimiter}) was proposed.
However, such a threshold only serves to artificially elevate the effective tensile strength of the material.
Furthermore, the calibration of this threshold is specific to the geometry and boundary conditions of the glacier. Although this fix may suppress the most visible symptoms, it does not resolve or explain the underlying issue.
To understand the cause of the issue, we next analyze the stress state immediately ahead of the crack tip.

\begin{figure}[h!]
	\centering
	\includegraphics[width=6.4in]{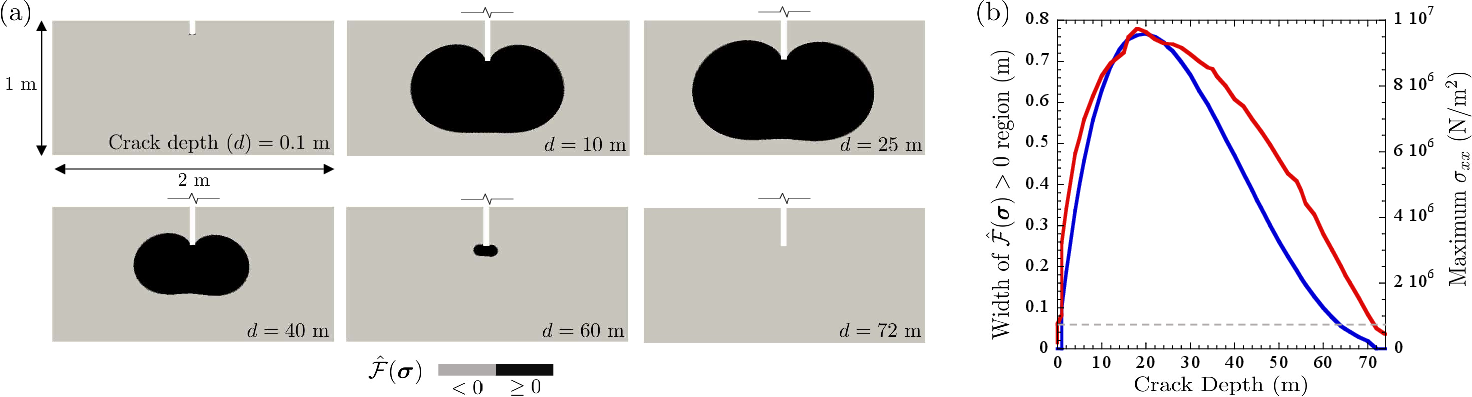}
	\caption{(a) Contour plots, over the undeformed configuration, of the regions of the glacier where the stress field exceeds ($\hat{\mathcal{F}}(\boldsymbol{\sigma}) \geq 0$) the strength surface of the ice. (b) Width of the region ahead of the crack tip where $\hat{\mathcal{F}}(\boldsymbol{\sigma}) \geq 0$, and maximum value of $\sigma_{\texttt{xx}}$, plotted as functions of crack depth measured from the top boundary.}\label{Fig4}
\end{figure}

We plot in Fig.~\ref{Fig4}(a) contour plots around the crack front of the regions where the strength surface $\hat{\mathcal{F}}(\boldsymbol{\sigma})$ is exceeded from purely elastic calculations of the stress field around the crack. The contour plots are shown at six increasing values of crack depths. We observe initially a clear increase in the width of the zone where the strength surface is violated with increasing crack depth.  This trend continues up to a depth of 18 m, after which the size of that zone starts decreasing and then ultimately vanishes. The results are quantified in Fig.~\ref{Fig4}(b), where the value of the maximum normal stress, $\sigma_{\texttt{xx}}$, just ahead of the crack is also reported. The maximum stress follows the same non-monotonic trend. 

At approximately a depth of 72 m, the maximum normal stress ahead of the crack becomes smaller in magnitude than the tensile strength of the ice. A crack would not be able to exceed this depth. This critical depth is not very sensitive to the value of tensile strength of ice in its realistic range of \{0.7, 2\} MPa. Interestingly, the critical depth for crack arrest obtained from the strength analysis matches approximately the value obtained from LEFM analysis \cite{sun2021poro}. Therefore, the critical crack depth is a strength-governed quantity and can be obtained without invoking fracture toughness or LEFM.

These results point to the fact that the glacier is an overdriven stress state in the presence of a stress concentrator. 
The width of the zone where strength is violated shows the region over which the stresses ahead of the crack tip are redistributed. If this region is small, as would be the case if the material strength is extremely high, a quasi-static analysis will suffice even if the crack growth is brutal. A wider zone means that a larger energy reservoir must be dissipated as the crack grows. In this case, a purely quasi-static analysis under fixed self–weight is not sufficient. 
Therefore, a dynamic analysis of the glacier is needed to model the abrupt, dynamic failure that occurs in glaciers from a crevasse.

\section{Analysis with a dynamic fracture model} \label{Sec: Dynamic}

\subsection{The phase-field theory of fracture nucleation and propagation}

Extending the equations of quasi-static elastic brittle fracture to dynamic fracture is not straightforward. Under dynamic loading, the \emph{apparent} strength surface and critical energy release rate may depend on the loading rate or crack velocity. Here, it is important to distinguish between apparent and intrinsic material properties, since material properties inferred from experiments often depend on the theoretical framework used for their interpretation. Moreover, the role of kinetic energy in the Griffith energetic competition between potential energy and fracture energy is not clear. 

A careful treatment of these issues would require a detailed reanalysis of dynamic fracture experiments and is beyond the scope of this work. We therefore extend the phase-field equations of quasi-static brittle fracture (\ref{phase-field-equations}) in the simplest possible manner: following Liu et al. \cite{liu2024dynamic}, we add the inertial term to the balance of linear momentum. Although this formulation may not be complete, Liu et al. showed that it agrees remarkably well with several dynamic fracture experiments.

There remains, however, the question of whether the mass density should be degraded, as discussed in the Introduction. Leaving the density undegraded allows a small amount of kinetic energy to persist in the cracked region, which can distort the highly degraded elements and lead to artificial widening of the crack band. We therefore first study the effect of degrading mass density through the following formulation.

Subject to the appropriate initial and boundary conditions, the displacement field $\mathbf{u}_k(\mathbf{X})=\mathbf{u}(\mathbf{X},t_k)$ and phase-field $v_k(\mathbf{X})=v(\mathbf{X},t_k)$ at any material point $\mathbf{X}\in\overline{\Omega}$ and discrete time $t_k\in\{0=t_0,t_1,\dots,t_m,$ $t_{m+1},\dots,$ $t_M=T\}$ are determined by the system of coupled partial differential equations 
\begin{equation}
\left\{\begin{array}{ll}
{\rm Div}\left[(v_{k}^2 + \eta)\dfrac{\partial W}{\partial \bfE}(\bfE(\bfu_{k}))\right]+(v_{k}^2 + \eta) \bfb
= (v_{k}^{b} +\eta^{\rho})\,\rho\,\ddot{\mathbf{u}}_{k},\\[10pt]
\dfrac{3}{4} \varepsilon \, \de \,  G_c \triangle v_{k}=2 v_{k} W(\bfE(\bfu_{k}))+c_\texttt{e}(\bfX,t_{k})- \dfrac{3}{8}  \dfrac{\de \, G_c}{\varepsilon}+\dfrac{\zeta}{2}\,p\!\left(v_{k-1},v_k\right), \\[10pt]
\end{array}\right. 
\label{phase-field-dynamics}
\end{equation}
where $\ddot{\mathbf{u}} = \frac{\partial^2 \mathbf{u}}{\partial t^2}$ stands for the acceleration field. The exponent $b$ controls the degradation of inertial term and takes the values $b=0$ or $b=2$ in this work. Other integer values could in principle be explored. The case $b=1$ gives an elastic wave speed $c=0$ at $v=0$ when $\eta=0$, while larger integer values may reduce the mass loss associated with density degradation. However, the results presented in the next section show that $b=2$ with $\eta^{\rho}=\eta=10^{-7}$ gives good results. We therefore restrict attention to these values.

In the following subsections, we compare the $b=0$ and $b=2$ formulations for several benchmark problems. Before doing so, we emphasize that, in this phase-field method, the mass loss associated with $b=2$ is only a regularization error of $O(\varepsilon)$. We also note that the elastic wave speed $c$ remains nonzero in the cracked region for $b=2$ and $\eta^{\rho}=\eta$. This is not problematic, since the relevant condition for the regularized crack to behave as a physical open crack is that its acoustic impedance, $Z=\rho c$, vanishes.

For the temporal discretization, we use the generalized-$\alpha$ method with $\alpha_m=0$, $\alpha_f=0.25$, $\beta=0.25(1+\alpha_f)^2$, and $\gamma=0.5+\alpha_f$, unless otherwise stated. The finite element implementation is discussed in Appendix B. All other aspects of the formulation and numerical implementation remain the same as in the quasi-static case.

\subsubsection{Wave propagation in a one-dimensional bar and a two-dimensional plate}

\begin{figure}[h!]
	\centering
	\includegraphics[width=6in]{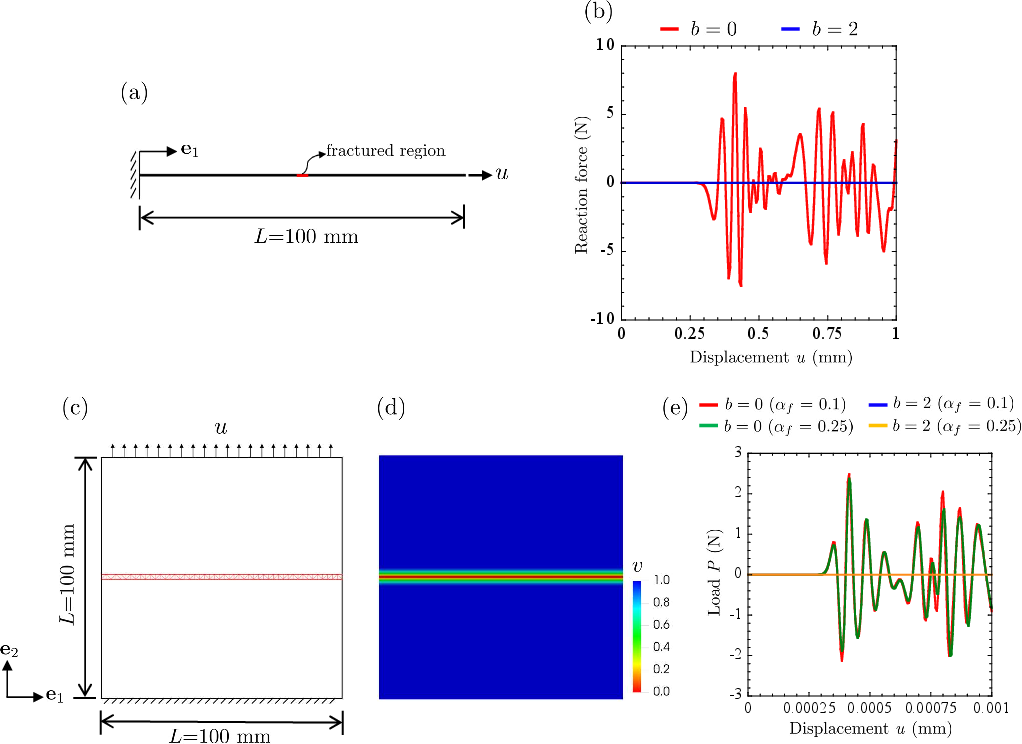}
	\caption{ (a) Schematic of the one-dimensional bar subjected to dynamic loading and containing a regularized crack at its center, with the phase-field profile given by \eqref{eq:degradation_function}. (b) Reaction force measured at the left end of the bar as a function of applied displacement for the $b=0$ case, where density is left undegraded, and the $b=2$ case, where density is degraded. (c) Schematic of the two-dimensional plate containing a central through-crack and subjected to dynamic loading. The phase-field crack is initialized by assigning $v=0$ to the two rows of nodes shown in the figure. (d) Developed phase-field crack profile. (e) Vertical reaction force measured at the bottom boundary as a function of applied displacement for the $b=0$ and $b=2$ cases, using two different values of the generalized-$\alpha$ parameter $\alpha_f$ in the temporal discretization.}
    \label{fig:wave}
\end{figure}

We begin with wave propagation in a one-dimensional elastic bar of length $L$ containing a regularized crack at its center, as shown in Fig.~\ref{fig:wave}(a). The governing equation for longitudinal wave propagation in an elastic medium is 
\begin{equation}
\rho' \frac{\partial^2 u}{\partial t^2} - \frac{\partial}{\partial x} \left( E' \frac{\partial u}{\partial x} \right) = 0. 
\label{eq:wave_equation}
\end{equation}
The fractured region is centered at $x_c=L/2$, where the elastic modulus and mass density are degraded as $E'=v^2E$ and $\rho'=v^b\rho$, respectively. The degradation function $v(x)$ is taken to be the optimal AT1 phase-field profile:
\begin{equation}
v(x) =
\begin{cases}
1 - \left(1 - \dfrac{|x - x_c|}{2\varepsilon}\right)^2, & \text{if } |x - x_c| \le 2\varepsilon, \\[8pt]
1, & \text{otherwise},
\end{cases}
\label{eq:degradation_function}
\end{equation}
The material properties are Young's modulus $E=25$ GPa, density $\rho=2450$ kg$/$m$^3$, and Poisson's ratio $\nu=0.19$. The bar length is $L=100$ mm and the domain is discretized using uniform elements of size $h=0.25$ mm. The regularization length is set to $\varepsilon=5h$, and the simulations are performed using a time step $\Delta t= 0.5 {\mu}$s.

A displacement $u$ is applied at the right end of the bar at a constant rate of 10 m/s. The reaction force at the left support, $X_1=0$, is plotted in Fig.~\ref{fig:wave}(b) as a function of the displacement applied at the right end, $X_1=L$, for the two cases of interest. For $b=0$, where density is not degraded, a small reaction force is registered at the left support. Although this force is very small, it shows that inertia retained in the damaged region, especially for finite $\varepsilon$, can lead to a small amount of force transmission across the regularized crack. In contrast, when the density is degraded with $b=2$, the reaction force is zero.

Consistent results are obtained for a similar analysis of a two-dimensional plate containing a through-crack, shown in Fig.~\ref{fig:wave}(c). In this example, rather than prescribing the full phase-field profile directly, we prescribe $v=0$ along a row of elements at $X_2=L/2$ and then solve the phase-field equation (\ref{phase-field-dynamics})$_2$ to obtain the optimal phase-field profile in the plate, as shown in Fig.~\ref{fig:wave}(d). 
The bottom boundary of the plate is then fully constrained in both the $x$ and $y$ directions, while the top boundary is subjected to a dynamic displacement $u$ at a constant rate of 10 mm/s. The same material properties and time-step size are used, and $\eta$ is set to $10^{-7}$.
The reaction force at the bottom support is plotted in Fig.~\ref{fig:wave}(e) as a function of the applied displacement. The results again show that a small force is transmitted through the cracked region for $b=0$, whereas degrading the density eliminates this spurious force transmission. We verified that these results are independent of both the temporal and spatial discretizations.

\subsubsection{Crack branching and mesh distortion in the Kalthoff-Winkler experiment}

\begin{figure}[h!]
	\centering
	\includegraphics[width=6.0in]{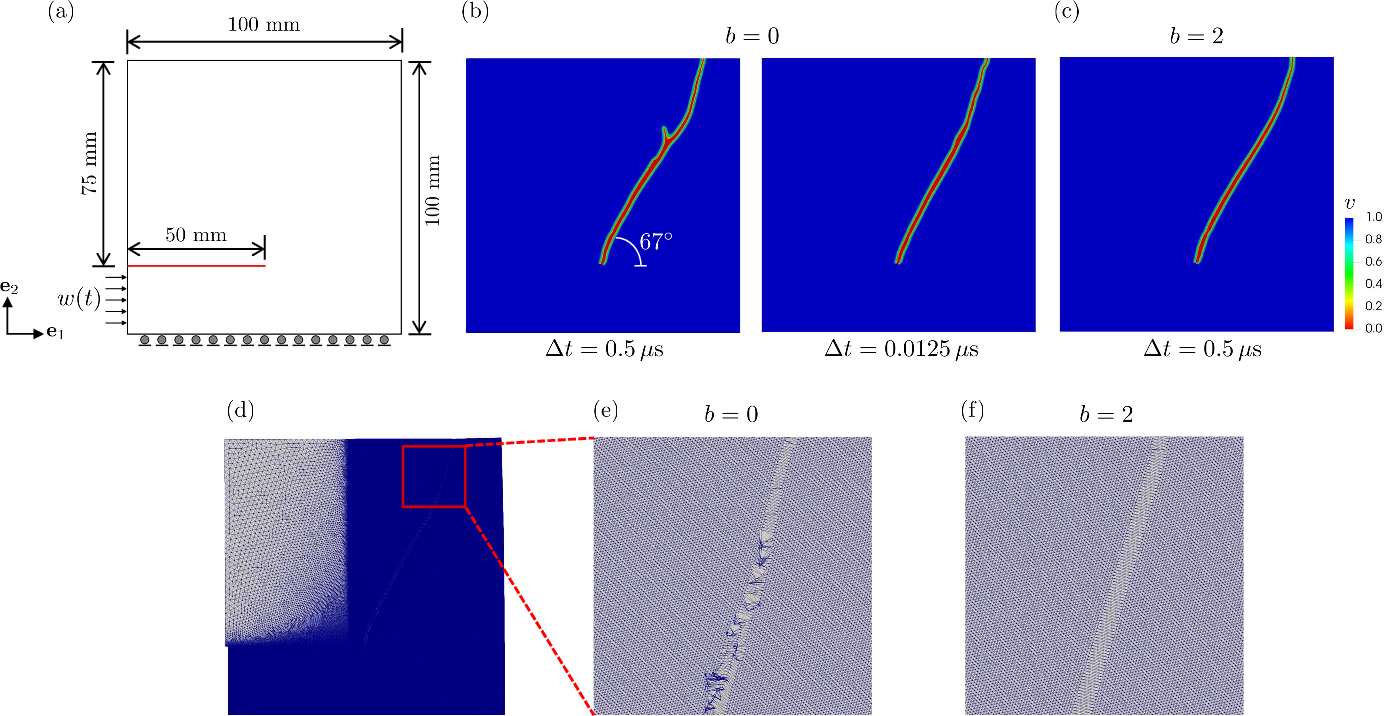}
	\caption{Simulations of the Kalthoff-Winkler test for the $b=0$ formulation, where density is left undegraded, and the $b=2$ formulation, where density is degraded. (a) Schematic and boundary conditions of the test. (b) Phase-field contours for the $b=0$ case using two time steps, $\Delta t=0.5~\mu$s and $\Delta t=0.0125~\mu$s. (c) Phase-field contour for the $b=2$ case with $\Delta t=0.5~\mu$s. (d) Finite element mesh used in the simulations, with mesh size $h=0.15625$ mm. Magnified views of the deformed mesh near the top boundary are shown for (e) $b=0$ and (f) $b=2$.}
    \label{fig:kalthoff}
\end{figure}

Next, we investigate the Kalthoff-Winkler experiment \cite{kalthoffwinkler1988}, a standard benchmark problem in dynamic fracture. A schematic of the geometry, dimensions, and boundary conditions is shown in Fig.~\ref{fig:kalthoff}(a). The geometry, material properties, and boundary conditions are taken from \cite{liu2024dynamic}. The material properties are Young's modulus $E=190$ GPa, density $\rho=8$ g$/$cm$^3$, Poisson's ratio $\nu=0.3$, tensile strength $\sigma_{\texttt{ts}}=1733$ MPa, compressive strength $\sigma_{\texttt{cs}}=5500$ MPa, and fracture toughness $G_c=22.2$ N/mm. The regularization length is set to $\varepsilon=0.75$ mm, and a uniform mesh size of $h=0.15$ mm is used in the regions through which the cracks propagate, as shown in Fig.~\ref{fig:kalthoff}(d). To approximate the projectile impact on the plate in the experiment, the following velocity $w(t)$ is prescribed on $X_1=0$ and $X_2\in[0,25]$ mm:
\[
w(t)
=
\begin{cases}
\dfrac{v_0}{t_0}\,t, & \text{if } t<t_0,\\[6pt]
v_0, & \text{if } t\geq t_0.
\end{cases}
\]
with $v_0 = 16.5$ m/s and $t_0= 1 \mu$s.

Fig.~\ref{fig:kalthoff}(b) shows the phase-field contour for the $b=0$ case, corresponding to undegraded density, using a time step $\Delta t=0.5~\mu$s and $\eta=10^{-4}$. The crack propagates upward at approximately $67^{\circ}$, consistent with experimental observations and previous numerical results. However, a small branch forms as the crack approaches the top boundary. This branching is spurious: it disappears when the time step is reduced to $\Delta t=0.0125~\mu$s. Since the generalized-$\alpha$ method is unconditionally stable for the parameters used here, this behavior is not a stability issue in the usual sense. The branching is also absent for $\Delta t=0.5~\mu$s when the generalized-$\alpha$ parameter $\alpha_f$ is set to zero, indicating sensitivity to the temporal discretization.  

We attribute this spurious branching to residual kinetic energy in the cracked region when the density is not degraded. The corresponding phase-field contours for the $b=2$ case are shown in Fig.~\ref{fig:kalthoff}(c) for $\Delta t=0.5~\mu$s and $\alpha_f=0.25$. In this case, the spurious branch does not appear, and the results are insensitive to both $\Delta t$ and $\alpha_f$.

In addition to spurious branching, the $b=0$ formulation requires a larger residual stiffness parameter $\eta$ to avoid severe mesh distortion in the cracked region. This distortion occurs because elements with nearly vanishing stiffness still carry inertia, which induces destabilizing oscillations of the highly degraded elements. Fig.~\ref{fig:kalthoff}(e) shows the $b=0$ result for $\eta=10^{-7}$, where excessive element distortion is observed. With the same value of $\eta$, the $b=2$ formulation shows no such distortion (Fig.~\ref{fig:kalthoff}(f)).

\subsubsection{Crack widening in the dynamic branching experiment}

\begin{figure}[h!]
	\centering
	\includegraphics[width=6.5in]{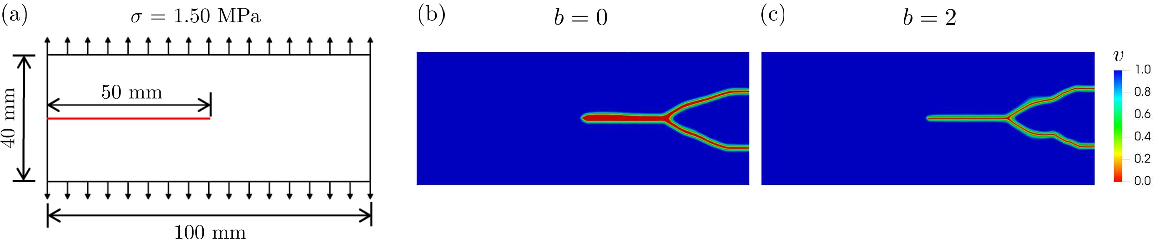}
	\caption{Simulations of the dynamic branching test for the $b=0$ formulation, where density is left undegraded, and the $b=2$ formulation, where density is degraded. (a) Schematic and boundary conditions of the test. Tensile stress of $\sigma= 1.50$ MPa is applied on the top and bottom boundaries suddenly. (b) Contour plots of the phase field for the $b=0$ case with residual stiffness parameter $\eta=10^{-4}$. (c) Contour plots of the phase field for the $b=2$ case with $\eta=10^{-7}$. }
    \label{fig:dynamic_branching}
\end{figure}

The last benchmark problem we examine is the dynamic crack branching problem, which we use to study the effect of the residual stiffness parameter $\eta$ on crack widening in the $b=0$ and $b=2$ formulations. This problem has been extensively studied both experimentally $\cite{RaviChandar1984_SteadyStateCrack, RaviChandar1984_CrackInitiationArrest, RaviChandar1984_MicrostructuralAspects, RaviChandar1998_DynamicFractureBrittle, RaviChandar1984_ExperimentalDynamicFracture, Ramulu1985_MechanicsCrackCurving}$ and numerically $\cite{Belytschko2003_DynamicCrackHyperbolicity, Borden2012_PhaseFieldDynamicBrittle, Geelen2019_PhaseField, Guo2019_KalthoffWinklerPeridynamics}$. The specimen geometry is shown in Fig.~\ref{fig:dynamic_branching}(a). A uniform tensile stress of $\sigma=1.5$ MPa is applied on the top and bottom boundaries during the first time step and is then held constant throughout the simulation.

The material properties are Young's modulus $E=32$ GPa, density $\rho=2450$ kg$/$m$^3$, Poisson's ratio $\nu=0.2$, tensile strength $\sigma_{\texttt{ts}}=3.08$ MPa, compressive strength $\sigma_{\texttt{cs}}=9.24$ MPa, and fracture toughness $G_c=0.003$ N/mm. The regularization length is set to $\varepsilon=0.625$ mm, and a uniform mesh size of $h=0.125$ mm is used. The time step is $\Delta t=5.0\times10^{-8}$ s.

When density is not degraded ($b=0$), the simulations show significant mesh distortion for very small values of the residual stiffness parameter, such as $\eta=10^{-7}$, similar to the behavior observed in the Kalthoff-Winkler test. Increasing $\eta$ to $10^{-4}$ removes the mesh distortion, but artificially widens the crack, as shown in Fig.~\ref{fig:dynamic_branching}(b). This widening occurs because residual kinetic energy remains in the cracked region, $0\leq v<1$. This kinetic energy prevents the phase field from attaining the optimal profile associated with minimization of the regularized surface energy. As a result, additional energy is dissipated in forming the crack, which can artificially reduce the crack speed.

The corresponding result for the degraded-density formulation ($b=2$) with $\eta=10^{-7}$ is shown in Fig.~\ref{fig:dynamic_branching}(c). In this case, mesh distortion is not observed for small deformations, even for very small values of $\eta$. Thus, the residual stiffness parameter can be chosen to be negligibly small without producing spurious crack widening.

\subsection{Dynamic fracture simulations of the glacier}

The observations from the benchmark studies in the previous subsection suggest the use of a degraded mass density in the dynamic fracture phase-field formulation (\ref{phase-field-dynamics}). This means that the elastic, gravitational, and inertial terms are consistently degraded with $v^2$ in the balance of linear momentum. We now perform dynamic simulations of the glacier containing a crevasse with this formulation.

\begin{figure}[h!]
	\centering
	\includegraphics[width=6.3in]{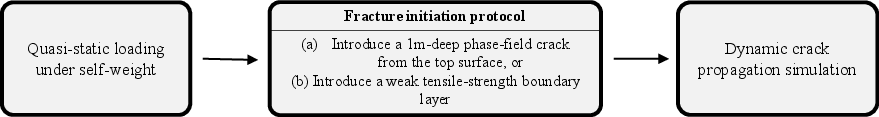}
	\caption{Schematic of the three-stage simulation procedure. A quasi-static simulation is first used to establish the gravitational pre-stress under self-weight. Fracture is then initiated either by prescribing a 1 m-deep surface crevasse at the centerline or by introducing a weak tensile-strength boundary layer. The resulting configuration is used as the initial state for the dynamic crack-propagation simulation.}\label{Fig5}
\end{figure}

The dynamic simulations are performed using a three–step procedure designed to separate the slow build–up of gravitational pre–stress from the fast brittle fracture growth. The three-step procedure is summarized in Fig.~\ref{Fig5} and proceeds as follows. First, we run a quasi-static simulation to load a pristine (no crevasses) glacier under self–weight to allow the displacement field, $\bfu_{\rm qs}(\bfX)$, to form. The tensile strength of the ice is significantly higher than the maximum principal stress that develops under self-weight, so no spontaneous crack formation occurs. Second, we trigger the onset of fracture at the upper surface using one of two equivalent seeding procedures: (i)  prescribing an initial, $1$~m deep phase–field crack by setting $v(\bfX,0)=0$ along a thin vertical region centered at $X_1=L/2$ and starting from the top boundary, or (ii) locally reducing the tensile strength $\sigma_{\texttt{ts}}$ to one-third it's actual value in a narrow region at the top boundary so that the same pre–stress state violates the strength surface and nucleates a crack naturally. The former is supposed to represent the formation of surface crevasses, and the latter represents a weak boundary layer due to loosely compacted snow or a melt-weakened band. Third, we start the dynamic solve using the prestressed quasi-static displacement field as the initial condition, $\bfu(\bfX,0)=\bfu_{\rm qs}(\bfX)$, with zero initial velocity and acceleration, and evolve the coupled system \eqref{phase-field-dynamics}.

\begin{figure}[h!]
	\centering
	\includegraphics[width=5.5in]{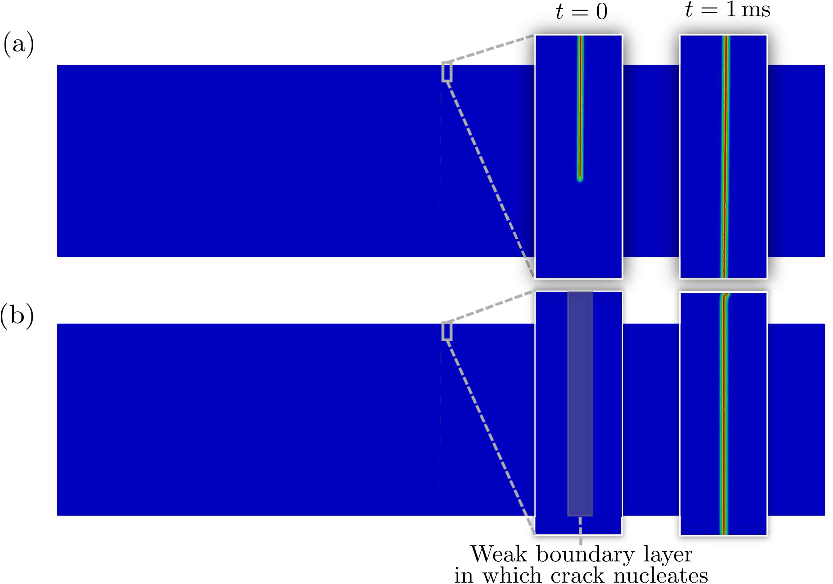}
	\caption{Contour plots of the phase field from dynamic fracture simulation of the glacier for two fracture initiation protocols: (a) A phase field crack of depth 1 m is introduced at $t=0$. The resulting crack growth is shown at a later time $t=1$ ms, and (b) A weak tensile-strength ($\sigma_{ts}=0.2$ MPa) boundary layer introduced in a 1 m deep, 6$\varepsilon$ wide region also results in a similar localized crack growth.}\label{Fig6}
\end{figure}

The geometry and material properties are the same as outlined in Section \ref{Sec: Problem} and a regularization length of $\varepsilon= 10$ mm is adopted. Fig.~\ref{Fig6}(a) shows the introduction of a small phase-field crack at $t=0$ when the dynamic computation is initiated.
Fig.~\ref{Fig6}(b) shows a result from the simulation at $t=5 \mu$s when the phase field has rapidly progressed downwards from the seeded crack. In contrast to the quasi-static results reported in Section \ref{Sec:quasistatic}, which showed a diffuse damage zone, the dynamic evolution produces a sharp, localized crack that propagates rapidly downward from the seeded crack. 

\begin{figure}[h!]
	\centering
	\includegraphics[width=3in]{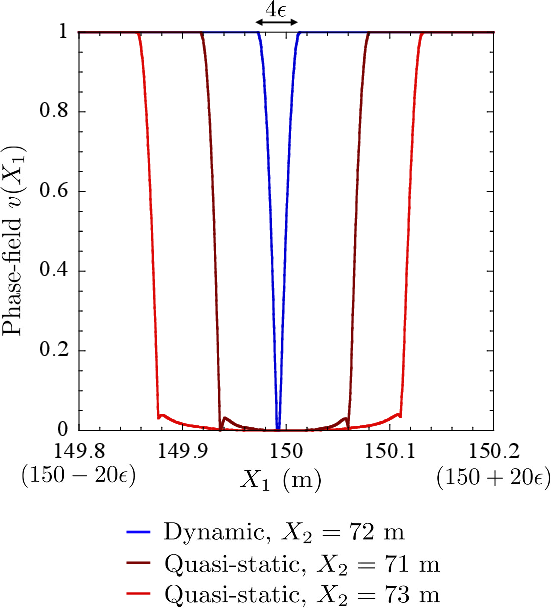}
	\caption{Phase-field profiles across a $40\varepsilon$-wide region centered on the crack path for simulations conducted with $\varepsilon= 10$ mm. The dynamic simulation produces the expected localized crack profile with a width of approximately $4\varepsilon$ at any depth (result shown for $X_2=72$ m), whereas the quasi-static simulations show substantial widening of the damaged region at different depths, $X_2=71$ m and $X_2=73$ m.}\label{Fig7}
\end{figure}

To demonstrate that the crack remains confined to a narrow diffusive band of thickness $4\, \varepsilon$ corresponding to the optimal width of the fractured region in the AT1 regularization, we plot the phase-field profile $v(x)$ at a depth of 72 m across a width of $40\varepsilon$ centered on the crack band, i.e., $x\in(150-20\varepsilon,\,150+20\varepsilon)$, as shown in Fig.~\ref{Fig7}. The result confirm that the phase field is localized. It is observed to independent of the vertical coordinate at which the cut is taken. For comparison, we also plot the phase-field profiles obtained from quasi-static computations at depths of 71 m and 73 m. They exhibit a widened damaged region of width significant greater than $4 \varepsilon$. They also show progressive widening as the crack advances. 

This comparison highlights the key point of this work: once the self–weight–driven stress state violates the strength surface by a significant margin, the subsequent evolution is an abrupt dynamic instability, and treating it dynamically prevents the diffuse, thickness–widening damage patterns characteristic of quasi-static evolutions observed in previous work.

The above behavior is also robust with respect to how the initial crack is seeded. Fig.~\ref{Fig6}(b) shows a result for the case where a upper boundary layer  with locally weakened $\sigma_{\texttt{ts}}$ is considered. The results show that  the resulting crack path and the through–thickness propagation dynamics are essentially identical to the case where the crack is seeded directly by prescribing $v$.

\begin{figure}[h!]
	\centering
	\includegraphics[width=6.3in]{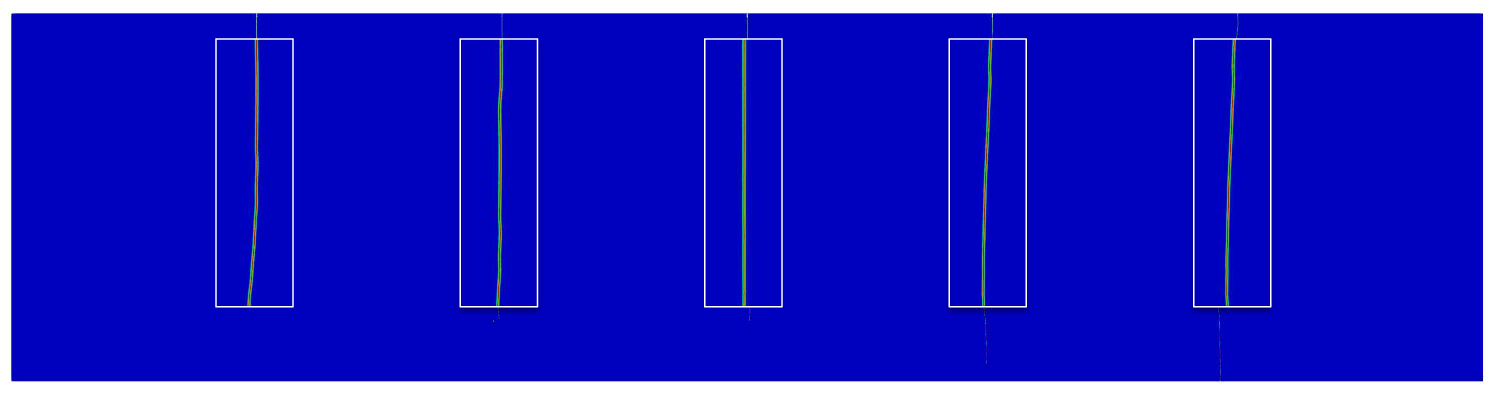}
	\caption{Multiple crevasse growth in a glacier simulated with the dynamic fracture formulation. Phase field contour shown for $t=1~\mu$s.}\label{Fig8}
\end{figure}

To analyze crevasse–to–crevasse interaction, we also perform simulations in which several surface crevasses are seeded along the upper boundary by prescribing $v(\bfX,0)=0$ at multiple locations ($X_1$ = 50 m, 100 m, 150 m, 200 m and 250 m). Under the same gravitational pre–stress obtained from the quasi\-static stage, the dynamic phase–field evolution predicts competitive growth: some crevasses arrest at finite depths while others propagate more rapidly, and neighboring cracks may shield one another spacing and the local stress concentration. A representative example is shown in Fig.~\ref{Fig8}.

Overall, the three-step simulation strategy provides a consistent computational route to simulate gravity–driven glacier fracture: the quasi-static stage establishes the pre–stress state, and the dynamic stage yields sharp, localized crack growth and realistic interaction of phase-field cracks.

\section{Final comments}\label{Sec: Final Comments}

We have shown that a purely quasi-static analysis of crevasse propagation under self-weight in grounded glaciers with the phase field model can lead to an unphysical widening of the crack band due to a large over-stressed zone that forms near the crack tip. We showed that this behavior can be remedied by using a dynamic fracture formulation. We extended previous phase-field models of dynamic fracture to consistently degrade the elastic, gravitational, and inertial terms in the equilibrium equation, and used this to show that a localized crack region of optimal width is obtained in simulations of glacier crevasse propagation.

In this work, motivated by the strong separation of time scales between slow viscous glacier flow and rapid fracture events, we idealized ice as a linear-elastic, isotropic, brittle solid. Furthermore, we adopted a two-dimensional plane strain analysis with idealized boundary conditions, omitting ocean hydrostatic pressure, tidal forcing, and evolving basal traction. 
This was done since the focus of this work was on presenting on the methodology of using the phase field method to analyze fracture in glaciers.

In future work, the methodology presented here can be extended to incorporate more realistic descriptions where needed. For example, coupling fracture with additional physics, such as the viscoplastic behavior of ice or thermomechanical effects, would require supplementing the present momentum balance and phase-field equations with the appropriate constitutive laws. Extensions to fluid-driven crack growth and poroelastic effects, which are needed to study the role of meltwater pressure in crevasse growth, can be developed along similar lines. The incompressible response of ice at slow strain rates can also be incorporated directly within this formulation, as demonstrated previously for elastomers \cite{KKLP24, dahal2025failure}. Finally, extension to three dimensions is straightforward and would enable the study of crevasse interactions and more realistic boundary conditions, such as hydrostatic ocean pressure acting on the glacier terminus. The use of adaptive mesh refinement algorithms will prove important there to mitigate the computational expense of three-dimensional phase-field fracture simulations \cite{gupta2024adaptive, nguyen2025adaptive}.
The methodology in the present work also provides a natural starting point for modeling fracture in floating ice shelves. In that setting, buoyancy, hydrostatic ocean pressure, and elastic bending interact with fracture, potentially giving rise to additional instabilities such as finger rafting \cite{vella2007finger}.

\section*{Acknowledgements}

\noindent {AK would like to acknowledge financial support from the Haythornthwaite Foundation Research Initiation Award. The computations reported here were conducted through research cyberinfrastructure resources and services provided by the Partnership for an Advanced Computing Environment (PACE) at the Georgia Institute of Technology. } 

\section*{Appendix A}

\begin{figure}[h!]
	\centering
	\includegraphics[width=6.3in]{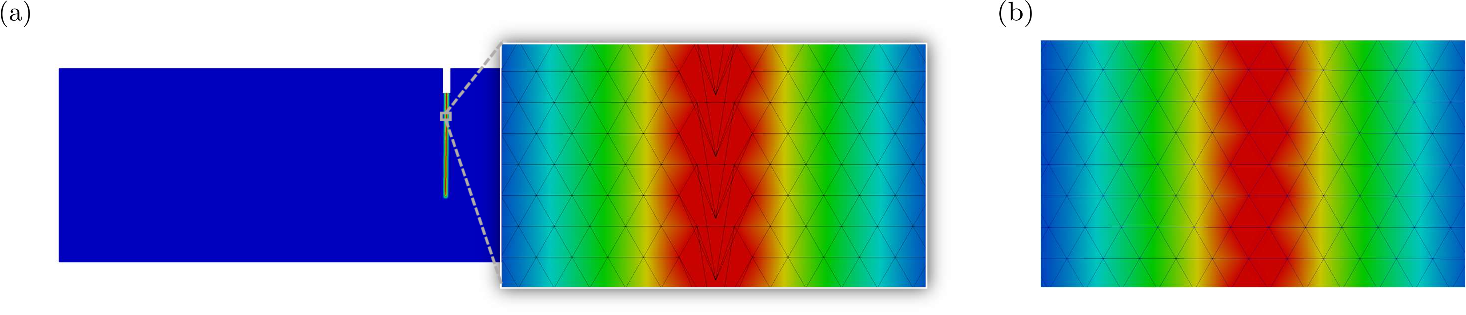}
	\caption{Deformed configuration of a narrow window around the crack tip of a gravity-loaded specimen with (a) phase field degradation not applied to the body force, and (b) degradation applied to the body force.}\label{Fig9}
\end{figure}
When modeling fracture under self-weight with regularized phase-field models, one must specify how the body force, or equivalently the mass density, is treated inside the fractured region. Two constitutive choices are possible: the density may be left undegraded, or it may be degraded with the phase field. If the density is kept constant, $\rho(v)\equiv\rho_0$, then the self-weight $\rho_0 \, \mathbf{g}$ continues to act inside the fractured region even as $v\to 0$, and the stiffness vanishes. The body force therefore continues to act on material that has effectively lost its ability to carry stress, producing an inconsistent local balance: the gravitational load remains finite while the internal resistance tends to zero. As a result, the displacement field becomes strongly ill-conditioned in the fractured region and may develop severe mesh distortion, as illustrated in Fig.~\ref{Fig9}(a). Although mesh distortion inside fully damaged regions is not necessarily problematic by itself, the presence of a small residual stiffness through a nonzero $\eta$ allows these regions to resist the body force weakly. This can artificially widen the fractured zone and slow crack propagation.

To avoid these artifacts, we degrade the body force consistently with $v^2$ throughout this work, using the same degradation applied to the stiffness. This choice produces a well-behaved displacement field in the fractured region, as shown in Fig.~\ref{Fig9}(b). Degrading the mass density does introduce a small loss of mass from the system; however, as discussed above, this loss is a regularization error of $O(\varepsilon)$.

\section*{Appendix B}

In this section, we present the finite element implementation of the phase field model for dynamic fracture as described by the equations ($\ref{phase-field-dynamics}$). The formulation uses the Galerkin finite element method for spatial discretization and the generalized-$\alpha$ method for temporal discretization. The discretization equations have been implemented in the open-source software FEniCSX.

\subsubsection*{{Weak form of the governing equations}}

We define the trial function spaces $\mathcal{S}_{t}$ and $\tilde{\mathcal{S}}_{t}$ for the displacement field and the phase-field respectively as 

\begin{equation}
\mathcal{S}_{t} = \{ \mathbf{u}(t) \in (H^{1}(\Omega))^{d} \mid \mathbf{u} = \bar{\mathbf{u}} \text{ on } \partial\mathrm{\Omega}_\mathcal{D} \},
\notag
\end{equation}
\begin{equation}
\tilde{\mathcal{S}}_{t} = \{ v(t) \in H^{1}(\Omega) \}.
\notag
\end{equation}
Similarly, we define the test function spaces as 
\begin{equation}
\mathcal{W} = \{ \mathbf{w} \in (H^{1}(\Omega))^{d} \mid \mathbf{w} = \mathbf{0} \text{ on } \partial\mathrm{\Omega}_\mathcal{D} \},
\notag
\end{equation}
\begin{equation}
\tilde{\mathcal{W}} = \{ q \in H^{1}(\Omega) \}.
\notag
\end{equation}
The continuous weak form of the force balance \eqref{phase-field-dynamics}$_1$ is given by :
\begin{equation}
\left\{\begin{array}{ll}
\text{Find } \mathbf{u}(t) \in \mathcal{S}(t), \\[8pt]
\int_{\Omega} v^2 \rho\, \ddot{\mathbf{u}} \cdot \mathbf{w}\,dx 
+ \int_{\Omega} v^2 \boldsymbol{\sigma}(\mathbf{u}) : \mathbf{E}(\mathbf{w})\,dx 
= \int_{\Omega} v^2 \mathbf{b} \cdot \mathbf{w}\,dx 
+ \int_{\partial \Omega} \mathbf{\bar{t}} \cdot \mathbf{w}\,ds,\quad \forall\, \mathbf{w}\in \mathcal{W}, \\[8pt]
\end{array}\right. 
\label{weakBVP-u-theory}
\end{equation}
for all $\mathbf{w} \in \mathcal{W}$ and where $\mathbf{\bar{t}}$ is the applied traction. Similarly, the weak form of the phase-field equation \eqref{phase-field-dynamics}$_2$ is obtained as
\begin{equation}
\left\{\begin{array}{ll}
\text{Find } v(t) \in \tilde{\mathcal{S}(t)}, \\[8pt]
\int_{\Omega} \dfrac{3}{4} \varepsilon \, \delta \,  G_c \nabla v \cdot \nabla q \, dx = \int_{\Omega} \left(2 v W(\mathbf{E}(\mathbf{u})) + c_e - \dfrac{3}{8}  \dfrac{\delta \, G_c}{\varepsilon} +\dfrac{\zeta}{2}\,p \right) q \, dx, \quad \forall q \in \tilde{\mathcal{W}}, \\[8pt]
\end{array}\right. 
\label{weakBVP-v-theory}
\end{equation} 

\vspace{0.1cm}
\subsubsection*{{Spatial discretizations of weak form}}

We consider paritions of the domain $\Omega$ into non-overlapping simplicial elements. We define conforming finite element spaces $\mathcal{S}_{t}^{h} \subset \mathcal{S}_{t}$, $\mathcal{W}^{h} \subset \mathcal{W}$, $\tilde{\mathcal{S}}^{h} \subset \tilde{\mathcal{S}}_{t}$, and $\tilde{\mathcal{W}}^{h} \subset \tilde{\mathcal{W}}$. The Galerkin form of the problem is stated as follows: find $\mathbf{u}^h(t) \in \mathcal{S}_{t}^{h}$ and $v^h(t) \in \tilde{\mathcal{S}}^{h}$ such that for all $\mathbf{w}^h \in \mathcal{W}^{h}$ and $q^h \in \tilde{\mathcal{W}}^{h}$:

\begin{equation} 
\int_{\Omega} (v^h)^2 \rho \ddot{\mathbf{u}}^h \cdot \mathbf{w}^h  dx + \int_{\Omega} (v^h)^2 \boldsymbol{\sigma}(\mathbf{u}^h) : \mathbf{E}(\mathbf{w}^h) dx = \int_{\Omega} (v^h)^2 \mathbf{b} \cdot \mathbf{w}^h  dx + \int_{\partial\Omega} \bar{\mathbf{t}} \cdot \mathbf{w}^h  ds ,
\end{equation}
and
\begin{equation} 
\int_{\Omega} \frac{3}{4} \varepsilon \delta {G}_c \nabla v^h \cdot \nabla q^h dx = \int_{\Omega} \left( 2 v^h W(\mathbf{E}(\mathbf{u}^h)) + c_e - \frac{3}{8} \frac{\delta {G}_c}{\varepsilon} +\dfrac{\zeta}{2}\,p \right) q^h  dx .
\end{equation}

The spatially discretized equations can be expressed in the following form:

$$M(\ddot{\mathbf{u}}^h, v^h, \mathbf{w}^h) + K_u(\mathbf{u}^h, v^h, \mathbf{w}^h) = F(v^h, \mathbf{w}^h)$$

$$A_v(v^h, q^h) = F_v(v^h, \mathbf{u}^h, q^h)$$

where:

$$M(\ddot{\mathbf{u}}, v, \mathbf{w}) = \int_{\Omega} v^2 \rho \ddot{\mathbf{u}} \cdot \mathbf{w} \, dx$$

$$K_u(\mathbf{u}, v, \mathbf{w}) = \int_{\Omega} v^2 \sigma(\mathbf{u}) : \mathbf{E}(\mathbf{w}) \, dx$$

$$F(v, \mathbf{w}) = \int_{\Omega} v^2 \mathbf{b} \cdot \mathbf{w} \, dx + \int_{\partial\Omega} \bar{\mathbf{t}} \cdot \mathbf{w} \, ds$$

$$A_v(v, q) = \int_{\Omega} \frac{3}{4} \varepsilon \delta {G}_c \nabla v \cdot \nabla q \, dx$$

$$F_v(v, \mathbf{u}, q) = \int_{\Omega} \left( 2 v W(\mathbf{E}(\mathbf{u})) + c_e - \frac{3}{8} \frac{\delta {G}_c}{\varepsilon} +\dfrac{\zeta}{2}\,p \right) q \, dx$$

\noindent
The approximation for the displacement field $\mathbf{u}^h$ and the phase-field $v^h$, and the test functions $\mathbf{w}^h$ and $q^h$ in the Galerkin method are given by:

$$\mathbf{u}^h(\mathbf{X}, t) = \sum_{i=1}^{N_e} N(\mathbf{X}) \tilde{\mathbf{u}}(t), \quad \mathbf{w}^h(\mathbf{x}) = \sum_{i=1}^{N_e} N(\mathbf{X}) \tilde{\mathbf{w}},$$

$$v^h(\mathbf{X}, t) = \sum_{i=1}^{N_e} N(\mathbf{X}) \tilde{v}(t), \quad q^h(\mathbf{x}) = \sum_{i=1}^{N_e} N(\mathbf{X}) \tilde{q},$$
where $N(\mathbf{X})$ are linear shape functions, $N_e$ represents number of nodes, $\tilde{\mathbf{u}}$ and  $\tilde{v}$ are the nodal displacements and phase-field variables, and $\tilde{\mathbf{w}}$ and  $\tilde{q}$ are the corresponding variations.
The resultant semidiscrete weak form of the displacement equation becomes 
$$\mathbf{M}(\tilde{v}) \ddot{\tilde{\mathbf{u}}} + \mathbf{K}_u(\tilde{v}) \tilde{\mathbf{u}} = \mathbf{F}(\tilde{v}),$$
where the components of the mass matrix $\mathbf{M}$, stiffness matrix $\mathbf{K}_u$, and external force vector $\mathbf{F}$ are:
$$\mathbf{M} = \int_{\Omega} (v^h)^2 \rho N N^T \, dx,$$

$$\mathbf{K}_{u} = \int_{\Omega} (v^h)^2 \mathbf{B}^T \mathbf{L} \mathbf{B} \, dx,$$

$$\mathbf{F} = \int_{\Omega} (v^h)^2 N \mathbf{b} \, dx + \int_{\partial\Omega} N \bar{\mathbf{t}} \, ds.$$
Here, $\mathbf{L}$ is the elasticity tensor and $\mathbf{B}$ is the strain-displacement matrix associated with node $i$. Note that the mass and stiffness matrices explicitly depend on the phase-field $\tilde{v}$.

\subsubsection*{{Time discretization using the generalized-$\alpha$ method}}

We discretize the time interval $[0,T]$ into $N{+}1$ steps 
$0=t_0 < t_1 < \cdots < t_N < t_{N+1}=T$  
with uniform time step $\Delta t = T/N$.  
The generalized-$\alpha$ method is used for temporal discretization.
It generalizes the classical Newmark-$\beta$ method and provides unconditional stability, controllable numerical dissipation, and second-order accuracy.

In this method, the spatially discretized dynamic equilibrium equation is evaluated at the intermediate time $t^{n+1-\alpha_f}$ (the tilde accent is dropped for convenience):
\begin{equation}
  \mathbf{M}\,\ddot{\mathbf{u}}^{\,n+1-\alpha_m}
  + \mathbf{K}\,\mathbf{u}^{\,n+1-\alpha_f}
  =
  \mathbf{F}(t^{\,n+1-\alpha_f}),
\end{equation}
where
\begin{equation}
  \mathbf{\ddot{u}}^{n+1-\alpha_m}_h
  =
  (1-\alpha_m)\,\mathbf{\ddot{u}}^{n+1}_h
  + \alpha_m\,\mathbf{\ddot{u}}^n_h
\end{equation}
and
\begin{equation}
  \mathbf{{u}}^{n+1-\alpha_f}_h
  =
  (1-\alpha_f)\,\mathbf{{u}}^{n+1}_h
  + \alpha_f\,\mathbf{{u}}^n_h.
\end{equation}
Using the Newmark-type update relations for displacement, velocity and acceleration:
\begin{equation}
\begin{aligned}
  \mathbf{u}^{n+1}
  &= 
  \mathbf{u}^{n}
  + \Delta t\,\dot{\mathbf{u}}^{n}
  + \frac{\Delta t^{2}}{2}\!\left[(1-2\beta)\,\ddot{\mathbf{u}}^{n}
  + 2\beta\,\ddot{\mathbf{u}}^{n+1}\right], \\[2mm]
  \dot{\mathbf{u}}^{n+1}
  &= 
  \dot{\mathbf{u}}^{n}
  + \Delta t\!\left[(1-\gamma)\,\ddot{\mathbf{u}}^{n}
  + \gamma\,\ddot{\mathbf{u}}^{n+1}\right] \\[2mm]
  \ddot{\mathbf{u}}^{n+1}
  &=
  \frac{1}{\beta\Delta t^{2}}
  \bigl(
    \mathbf{u}^{n+1}
    - \mathbf{u}^{n}
    - \Delta t\,\dot{\mathbf{u}}^{n}
  \bigr)
  - \frac{1-2\beta}{2\beta}\,\ddot{\mathbf{u}}^{n},
\end{aligned}
\end{equation}
one can solve for the acceleration $\mathbf{\ddot{u}}_h$ at time $t^{n+1-\alpha_m}$ as
\begin{equation}
  \mathbf{\ddot{u}}^{n+1-\alpha_m}_h
  =
  m_1 \mathbf{u}^{n+1}_h
  - m_1 \mathbf{u}^n_h
  - m_2 \mathbf{\dot{u}}^n_h
  + m_3 \mathbf{\ddot{u}}^n_h .
\end{equation}

\noindent
where $m_1=({1-\alpha_m})/({\beta\Delta t^2})$, $m_2=({1-\alpha_m})/({\beta\Delta t})$, and $m_3=({1-\alpha_m-2\beta})/({2\beta})$.

Substituting the above expression into the equilibrium equation and gathering all terms involving the unknown $\mathbf{u}^{n+1}_h$ on the left hand side yields the final linear system
\begin{equation}
  \mathbf{\bar{K}}\,\mathbf{u}^{n+1}_h
  =
  \mathbf{F}(t^{\,n+1-\alpha_f})
  - \alpha_f\,\mathbf{K}\,\mathbf{u}^{n}_h
  + \mathbf{M}\!\left(m_1\mathbf{u}^{n}_h + m_2 \dot{\mathbf{u}}^{n}_h - m_3 \ddot{\mathbf{u}}^{n}_h\right),
\end{equation}
with the effective stiffness matrix
\begin{equation}
  \mathbf{\bar{K}}
  =
  (1-\alpha_f)\,\mathbf{K}
  + m_1\,\mathbf{M}.
\end{equation}

\bibliographystyle{elsarticle-num-names}
\bibliography{new}

\begin{thebibliography}{66}
\expandafter\ifx\csname natexlab\endcsname\relax\def\natexlab#1{#1}\fi
\providecommand{\url}[1]{\texttt{#1}}
\providecommand{\href}[2]{#2}
\providecommand{\path}[1]{#1}
\providecommand{\DOIprefix}{doi:}
\providecommand{\ArXivprefix}{arXiv:}
\providecommand{\URLprefix}{URL: }
\providecommand{\Pubmedprefix}{pmid:}
\providecommand{\doi}[1]{\href{http://dx.doi.org/#1}{\path{#1}}}
\providecommand{\Pubmed}[1]{\href{pmid:#1}{\path{#1}}}
\providecommand{\bibinfo}[2]{#2}
\ifx\xfnm\relax \def\xfnm[#1]{\unskip,\space#1}\fi
\bibitem[{Meier et~al.(2007)Meier, Dyurgerov, Rick, O'neel, Pfeffer, Anderson, Anderson, and Glazovsky}]{meier2007glaciers}
\bibinfo{author}{M.~F. Meier}, \bibinfo{author}{M.~B. Dyurgerov}, \bibinfo{author}{U.~K. Rick}, \bibinfo{author}{S.~O'neel}, \bibinfo{author}{W.~T. Pfeffer}, \bibinfo{author}{R.~S. Anderson}, \bibinfo{author}{S.~P. Anderson}, \bibinfo{author}{A.~F. Glazovsky},
\newblock \bibinfo{title}{Glaciers dominate eustatic sea-level rise in the 21st century},
\newblock \bibinfo{journal}{Science} \bibinfo{volume}{317} (\bibinfo{year}{2007}) \bibinfo{pages}{1064--1067}.
\bibitem[{Scambos et~al.(2000)Scambos, Hulbe, Fahnestock, and Bohlander}]{scambos2000link}
\bibinfo{author}{T.~A. Scambos}, \bibinfo{author}{C.~Hulbe}, \bibinfo{author}{M.~Fahnestock}, \bibinfo{author}{J.~Bohlander},
\newblock \bibinfo{title}{The link between climate warming and break-up of ice shelves in the antarctic peninsula},
\newblock \bibinfo{journal}{Journal of Glaciology} \bibinfo{volume}{46} (\bibinfo{year}{2000}) \bibinfo{pages}{516--530}.
\bibitem[{Bassis and Jacobs(2013)}]{bassis2013diverse}
\bibinfo{author}{J.~N. Bassis}, \bibinfo{author}{S.~Jacobs},
\newblock \bibinfo{title}{Diverse calving patterns linked to glacier geometry},
\newblock \bibinfo{journal}{Nature Geoscience} \bibinfo{volume}{6} (\bibinfo{year}{2013}) \bibinfo{pages}{833--836}.
\bibitem[{Benn et~al.(2017)Benn, Cowton, Todd, and Luckman}]{benn2017glacier}
\bibinfo{author}{D.~I. Benn}, \bibinfo{author}{T.~Cowton}, \bibinfo{author}{J.~Todd}, \bibinfo{author}{A.~Luckman},
\newblock \bibinfo{title}{Glacier calving in greenland},
\newblock \bibinfo{journal}{Current Climate Change Reports} \bibinfo{volume}{3} (\bibinfo{year}{2017}) \bibinfo{pages}{282--290}.
\bibitem[{Vaughan(1993)}]{vaughan1993relating}
\bibinfo{author}{D.~G. Vaughan},
\newblock \bibinfo{title}{Relating the occurrence of crevasses to surface strain rates},
\newblock \bibinfo{journal}{Journal of Glaciology} \bibinfo{volume}{39} (\bibinfo{year}{1993}) \bibinfo{pages}{255--266}.
\bibitem[{Van~der Veen(1998)}]{van1998fracture}
\bibinfo{author}{C.~Van~der Veen},
\newblock \bibinfo{title}{Fracture mechanics approach to penetration of surface crevasses on glaciers},
\newblock \bibinfo{journal}{Cold Regions Science and Technology} \bibinfo{volume}{27} (\bibinfo{year}{1998}) \bibinfo{pages}{31--47}.
\bibitem[{Kingslake et~al.(2017)Kingslake, Ely, Das, and Bell}]{kingslake2017widespread}
\bibinfo{author}{J.~Kingslake}, \bibinfo{author}{J.~C. Ely}, \bibinfo{author}{I.~Das}, \bibinfo{author}{R.~E. Bell},
\newblock \bibinfo{title}{Widespread movement of meltwater onto and across antarctic ice shelves},
\newblock \bibinfo{journal}{Nature} \bibinfo{volume}{544} (\bibinfo{year}{2017}) \bibinfo{pages}{349--352}.
\bibitem[{Pralong and Funk(2005)}]{pralong2005dynamic}
\bibinfo{author}{A.~Pralong}, \bibinfo{author}{M.~Funk},
\newblock \bibinfo{title}{Dynamic damage model of crevasse opening and application to glacier calving},
\newblock \bibinfo{journal}{Journal of Geophysical Research: Solid Earth} \bibinfo{volume}{110} (\bibinfo{year}{2005}).
\bibitem[{Weiss(2004)}]{weiss2004subcritical}
\bibinfo{author}{J.~Weiss},
\newblock \bibinfo{title}{Subcritical crack propagation as a mechanism of crevasse formation and iceberg calving},
\newblock \bibinfo{journal}{Journal of Glaciology} \bibinfo{volume}{50} (\bibinfo{year}{2004}) \bibinfo{pages}{109--115}.
\bibitem[{Bourdin et~al.(2000)Bourdin, Francfort, and Marigo}]{bourdin2000numerical}
\bibinfo{author}{B.~Bourdin}, \bibinfo{author}{G.~A. Francfort}, \bibinfo{author}{J.-J. Marigo},
\newblock \bibinfo{title}{Numerical experiments in revisited brittle fracture},
\newblock \bibinfo{journal}{Journal of the Mechanics and Physics of Solids} \bibinfo{volume}{48} (\bibinfo{year}{2000}) \bibinfo{pages}{797--826}.
\bibitem[{Francfort and Marigo(1998)}]{Francfort98}
\bibinfo{author}{G.~A. Francfort}, \bibinfo{author}{J.-J. Marigo},
\newblock \bibinfo{title}{Revisiting brittle fracture as an energy minimization problem},
\newblock \bibinfo{journal}{Journal of the Mechanics and Physics of Solids} \bibinfo{volume}{46} (\bibinfo{year}{1998}) \bibinfo{pages}{1319--1342}.
\bibitem[{Sun et~al.(2021)Sun, Duddu, and Hirshikesh}]{sun2021poro}
\bibinfo{author}{X.~Sun}, \bibinfo{author}{R.~Duddu}, \bibinfo{author}{Hirshikesh},
\newblock \bibinfo{title}{A poro-damage phase field model for hydrofracturing of glacier crevasses},
\newblock \bibinfo{journal}{Extreme Mechanics Letters} \bibinfo{volume}{45} (\bibinfo{year}{2021}) \bibinfo{pages}{101277}.
\bibitem[{Miehe et~al.(2015)Miehe, Schaenzel, and Ulmer}]{miehe2015}
\bibinfo{author}{C.~Miehe}, \bibinfo{author}{L.-M. Schaenzel}, \bibinfo{author}{H.~Ulmer},
\newblock \bibinfo{title}{Phase field modeling of fracture in multi-physics problems. part i. balance of crack surface and failure criteria for brittle crack propagation in thermo-elastic solids},
\newblock \bibinfo{journal}{Computer Methods in Applied Mechanics and Engineering} \bibinfo{volume}{294} (\bibinfo{year}{2015}) \bibinfo{pages}{449--485}.
\bibitem[{Clayton et~al.(2022)Clayton, Duddu, Siegert, and Martinez-Paneda}]{clayton2022stress}
\bibinfo{author}{T.~Clayton}, \bibinfo{author}{R.~Duddu}, \bibinfo{author}{M.~Siegert}, \bibinfo{author}{E.~Martinez-Paneda},
\newblock \bibinfo{title}{A stress-based poro-damage phase field model for hydrofracturing of creeping glaciers and ice shelves},
\newblock \bibinfo{journal}{Engineering Fracture Mechanics} \bibinfo{volume}{272} (\bibinfo{year}{2022}) \bibinfo{pages}{108693}.
\bibitem[{Sondershaus et~al.(2023)Sondershaus, Humbert, and M{\"u}ller}]{sondershaus2023phase}
\bibinfo{author}{R.~Sondershaus}, \bibinfo{author}{A.~Humbert}, \bibinfo{author}{R.~M{\"u}ller},
\newblock \bibinfo{title}{A phase field model for fractures in ice shelves},
\newblock \bibinfo{journal}{PAMM} \bibinfo{volume}{22} (\bibinfo{year}{2023}) \bibinfo{pages}{e202200256}.
\bibitem[{Khayaz et~al.(2025)Khayaz, Dahal, and Kumar}]{khayaz2025comparison}
\bibinfo{author}{U.~Khayaz}, \bibinfo{author}{A.~Dahal}, \bibinfo{author}{A.~Kumar},
\newblock \bibinfo{title}{A comparison of phase field models of brittle fracture incorporating strength: I—mixed-mode loading},
\newblock \bibinfo{journal}{Engineering Fracture Mechanics}  (\bibinfo{year}{2025}) \bibinfo{pages}{111679}.
\bibitem[{Kumar et~al.(2018)Kumar, Francfort, and Lopez-Pamies}]{KFLP18}
\bibinfo{author}{A.~Kumar}, \bibinfo{author}{G.~A. Francfort}, \bibinfo{author}{O.~Lopez-Pamies},
\newblock \bibinfo{title}{Fracture and healing of elastomers: A phase-transition theory and numerical implementation},
\newblock \bibinfo{journal}{Journal of the Mechanics and Physics of Solids} \bibinfo{volume}{112} (\bibinfo{year}{2018}) \bibinfo{pages}{523--551}.
\bibitem[{Kumar et~al.(2020{\natexlab{a}})Kumar, Bourdin, Francfort, and Lopez-Pamies}]{KBFLP20}
\bibinfo{author}{A.~Kumar}, \bibinfo{author}{B.~Bourdin}, \bibinfo{author}{G.~A. Francfort}, \bibinfo{author}{O.~Lopez-Pamies},
\newblock \bibinfo{title}{Revisiting nucleation in the phase-field approach to brittle fracture},
\newblock \bibinfo{journal}{Journal of the Mechanics and Physics of Solids} \bibinfo{volume}{142} (\bibinfo{year}{2020}{\natexlab{a}}) \bibinfo{pages}{104027}.
\bibitem[{Kumar et~al.(2020{\natexlab{b}})Kumar, Bourdin, Francfort, and Lopez-Pamies}]{kumar2020revisiting}
\bibinfo{author}{A.~Kumar}, \bibinfo{author}{B.~Bourdin}, \bibinfo{author}{G.~A. Francfort}, \bibinfo{author}{O.~Lopez-Pamies},
\newblock \bibinfo{title}{Revisiting nucleation in the phase-field approach to brittle fracture},
\newblock \bibinfo{journal}{Journal of the Mechanics and Physics of Solids} \bibinfo{volume}{142} (\bibinfo{year}{2020}{\natexlab{b}}) \bibinfo{pages}{104027}.
\bibitem[{Kumar and Lopez-Pamies(2021)}]{KLP21}
\bibinfo{author}{A.~Kumar}, \bibinfo{author}{O.~Lopez-Pamies},
\newblock \bibinfo{title}{The poker-chip experiments of gent and lindley (1959) explained},
\newblock \bibinfo{journal}{Journal of the Mechanics and Physics of Solids} \bibinfo{volume}{150} (\bibinfo{year}{2021}) \bibinfo{pages}{104359}.
\bibitem[{Kumar et~al.(2022)Kumar, Ravi-Chandar, and Lopez-Pamies}]{KRLP22}
\bibinfo{author}{A.~Kumar}, \bibinfo{author}{K.~Ravi-Chandar}, \bibinfo{author}{O.~Lopez-Pamies},
\newblock \bibinfo{title}{The revisited phase-field approach to brittle fracture: application to indentation and notch problems},
\newblock \bibinfo{journal}{International Journal of Fracture} \bibinfo{volume}{237} (\bibinfo{year}{2022}) \bibinfo{pages}{83--100}.
\bibitem[{Kumar et~al.(2024)Kumar, Liu, Dolbow, and Lopez-Pamies}]{KLDLP23}
\bibinfo{author}{A.~Kumar}, \bibinfo{author}{Y.~Liu}, \bibinfo{author}{J.~E. Dolbow}, \bibinfo{author}{O.~Lopez-Pamies},
\newblock \bibinfo{title}{The strength of the brazilian fracture test},
\newblock \bibinfo{journal}{Journal of the Mechanics and Physics of Solids} \bibinfo{volume}{182} (\bibinfo{year}{2024}) \bibinfo{pages}{105473}.
\bibitem[{Liu and Kumar(2025)}]{LK24}
\bibinfo{author}{C.~Liu}, \bibinfo{author}{A.~Kumar},
\newblock \bibinfo{title}{Emergence of tension–compression asymmetry from a complete phase-field approach to brittle fracture},
\newblock \bibinfo{journal}{International Journal of Solids and Structures} \bibinfo{volume}{309} (\bibinfo{year}{2025}) \bibinfo{pages}{113170}.
\bibitem[{Kamarei et~al.(2024)Kamarei, Kumar, and Lopez-Pamies}]{KKLP24}
\bibinfo{author}{F.~Kamarei}, \bibinfo{author}{A.~Kumar}, \bibinfo{author}{O.~Lopez-Pamies},
\newblock \bibinfo{title}{The poker-chip experiments of synthetic elastomers explained},
\newblock \bibinfo{journal}{Journal of the Mechanics and Physics of Solids}  (\bibinfo{year}{2024}) \bibinfo{pages}{105683}.
\bibitem[{Kamarei et~al.(2026)Kamarei, Zeng, Dolbow, and Lopez-Pamies}]{kamarei2026nine}
\bibinfo{author}{F.~Kamarei}, \bibinfo{author}{B.~Zeng}, \bibinfo{author}{J.~E. Dolbow}, \bibinfo{author}{O.~Lopez-Pamies},
\newblock \bibinfo{title}{Nine circles of elastic brittle fracture: A series of challenge problems to assess fracture models},
\newblock \bibinfo{journal}{Computer Methods in Applied Mechanics and Engineering} \bibinfo{volume}{448} (\bibinfo{year}{2026}) \bibinfo{pages}{118449}.
\bibitem[{Ward and Kumar(2026)}]{WK2025}
\bibinfo{author}{O.~Ward}, \bibinfo{author}{A.~Kumar},
\newblock \bibinfo{title}{Why planar cracks fragment into echelon cracks},
\newblock \bibinfo{journal}{Extreme Mechanics Letters}  (\bibinfo{year}{2026}) \bibinfo{pages}{102504}.
\bibitem[{Gr{\"a}ff et~al.(2025)Gr{\"a}ff, Lipovsky, Vieli, Dachauer, Jackson, Farinotti, Schmale, Ampuero, Berg, Dannowski et~al.}]{graff2025calving}
\bibinfo{author}{D.~Gr{\"a}ff}, \bibinfo{author}{B.~P. Lipovsky}, \bibinfo{author}{A.~Vieli}, \bibinfo{author}{A.~Dachauer}, \bibinfo{author}{R.~Jackson}, \bibinfo{author}{D.~Farinotti}, \bibinfo{author}{J.~Schmale}, \bibinfo{author}{J.-P. Ampuero}, \bibinfo{author}{E.~Berg}, \bibinfo{author}{A.~Dannowski}, et~al.,
\newblock \bibinfo{title}{Calving-driven fjord dynamics resolved by seafloor fibre sensing},
\newblock \bibinfo{journal}{Nature} \bibinfo{volume}{644} (\bibinfo{year}{2025}) \bibinfo{pages}{404--412}.
\bibitem[{Kooij et~al.(2021)Kooij, van Dalen, Molinari, and Bonn}]{kooij2021prince}
\bibinfo{author}{S.~Kooij}, \bibinfo{author}{G.~van Dalen}, \bibinfo{author}{J.-F. Molinari}, \bibinfo{author}{D.~Bonn},
\newblock \bibinfo{title}{Explosive fragmentation of prince rupert’s drops leads to well-defined fragment sizes},
\newblock \bibinfo{journal}{Nature communications} \bibinfo{volume}{12} (\bibinfo{year}{2021}) \bibinfo{pages}{2521}.
\bibitem[{Liu et~al.(2024)Liu, Lopez-Pamies, and Dolbow}]{liu2024dynamic}
\bibinfo{author}{Y.~Liu}, \bibinfo{author}{O.~Lopez-Pamies}, \bibinfo{author}{J.~E. Dolbow},
\newblock \bibinfo{title}{On the effects of material strength in dynamic fracture: A phase-field study},
\newblock \bibinfo{journal}{arXiv preprint arXiv:2411.16393}  (\bibinfo{year}{2024}).
\bibitem[{Bourdin et~al.(2011)Bourdin, Larsen, and Richardson}]{Bourdin2011_TimeDiscreteDynamic}
\bibinfo{author}{B.~Bourdin}, \bibinfo{author}{C.~J. Larsen}, \bibinfo{author}{C.~L. Richardson},
\newblock \bibinfo{title}{A time-discrete model for dynamic fracture based on crack regularization},
\newblock \bibinfo{journal}{International Journal of Fracture} \bibinfo{volume}{168} (\bibinfo{year}{2011}) \bibinfo{pages}{133--143}. \DOIprefix\doi{10.1007/s10704-010-9562-x}.
\bibitem[{Borden et~al.(2012)Borden, Verhoosel, Scott, Hughes, and Landis}]{Borden2012_PhaseFieldDynamicBrittle}
\bibinfo{author}{M.~J. Borden}, \bibinfo{author}{C.~V. Verhoosel}, \bibinfo{author}{M.~A. Scott}, \bibinfo{author}{T.~J. Hughes}, \bibinfo{author}{C.~M. Landis},
\newblock \bibinfo{title}{A phase-field description of dynamic brittle fracture},
\newblock \bibinfo{journal}{Computer Methods in Applied Mechanics and Engineering} \bibinfo{volume}{217--220} (\bibinfo{year}{2012}) \bibinfo{pages}{77--95}. \DOIprefix\doi{10.1016/j.cma.2012.01.008}.
\bibitem[{Hofacker and Miehe(2012)}]{Hofacker2012_ContinuumPhaseField}
\bibinfo{author}{M.~Hofacker}, \bibinfo{author}{C.~Miehe},
\newblock \bibinfo{title}{Continuum phase field modeling of dynamic fracture: variational principles and staggered fe implementation},
\newblock \bibinfo{journal}{International Journal of Fracture} \bibinfo{volume}{178} (\bibinfo{year}{2012}) \bibinfo{pages}{113--129}. \DOIprefix\doi{10.1007/s10704-012-9753-8}.
\bibitem[{Bleyer et~al.(2017)Bleyer, Roux-Langlois, and Molinari}]{Bleyer2017_DynamicCrackVariational}
\bibinfo{author}{J.~Bleyer}, \bibinfo{author}{C.~Roux-Langlois}, \bibinfo{author}{J.-F. Molinari},
\newblock \bibinfo{title}{Dynamic crack propagation with a variational phase-field model: limiting speed, crack branching and velocity-toughening mechanisms},
\newblock \bibinfo{journal}{International Journal of Fracture} \bibinfo{volume}{204} (\bibinfo{year}{2017}) \bibinfo{pages}{79--100}. \DOIprefix\doi{10.1007/s10704-016-0163-1}.
\bibitem[{Nguyen and Wu(2018)}]{Nguyen2018_PhaseFieldCohesive}
\bibinfo{author}{V.~P. Nguyen}, \bibinfo{author}{J.-Y. Wu},
\newblock \bibinfo{title}{Modeling dynamic fracture of solids with a phase-field regularized cohesive zone model},
\newblock \bibinfo{journal}{Computer Methods in Applied Mechanics and Engineering} \bibinfo{volume}{340} (\bibinfo{year}{2018}) \bibinfo{pages}{1000--1022}. \DOIprefix\doi{10.1016/j.cma.2018.06.010}.
\bibitem[{Geelen et~al.(2019)Geelen, Liu, Hu, Tupek, and Dolbow}]{Geelen2019_PhaseField}
\bibinfo{author}{R.~J. Geelen}, \bibinfo{author}{Y.~Liu}, \bibinfo{author}{T.~Hu}, \bibinfo{author}{M.~R. Tupek}, \bibinfo{author}{J.~E. Dolbow},
\newblock \bibinfo{title}{A phase-field formulation for dynamic cohesive fracture},
\newblock \bibinfo{journal}{Computer Methods in Applied Mechanics and Engineering} \bibinfo{volume}{348} (\bibinfo{year}{2019}) \bibinfo{pages}{680--711}. \DOIprefix\doi{10.1016/j.cma.2019.01.026}.
\bibitem[{Mandal et~al.(2020)Mandal, Nguyen, and Wu}]{Mandal2020_EvaluationVariational}
\bibinfo{author}{T.~K. Mandal}, \bibinfo{author}{V.~P. Nguyen}, \bibinfo{author}{J.-Y. Wu},
\newblock \bibinfo{title}{Evaluation of variational phase-field models for dynamic brittle fracture},
\newblock \bibinfo{journal}{Engineering Fracture Mechanics} \bibinfo{volume}{235} (\bibinfo{year}{2020}) \bibinfo{pages}{107169}. \DOIprefix\doi{10.1016/j.engfracmech.2020.107169}.
\bibitem[{Li et~al.(2016)Li, Marigo, Guilbaud, and Potapov}]{marigo2016dynamics}
\bibinfo{author}{T.~Li}, \bibinfo{author}{J.-J. Marigo}, \bibinfo{author}{D.~Guilbaud}, \bibinfo{author}{S.~Potapov},
\newblock \bibinfo{title}{Gradient damage modeling of brittle fracture in an explicit dynamics context},
\newblock \bibinfo{journal}{International Journal for Numerical Methods in Engineering} \bibinfo{volume}{108} (\bibinfo{year}{2016}) \bibinfo{pages}{1381--1405}.
\bibitem[{Montagnat and Duval(2004)}]{montagnat2004viscoplastic}
\bibinfo{author}{M.~Montagnat}, \bibinfo{author}{P.~Duval},
\newblock \bibinfo{title}{The viscoplastic behaviour of ice in polar ice sheets: experimental results and modelling},
\newblock \bibinfo{journal}{Comptes Rendus Physique} \bibinfo{volume}{5} (\bibinfo{year}{2004}) \bibinfo{pages}{699--708}.
\bibitem[{Deng et~al.(2020)Deng, Feng, Tan, and Hu}]{deng2020experimental}
\bibinfo{author}{K.~Deng}, \bibinfo{author}{X.~Feng}, \bibinfo{author}{X.~Tan}, \bibinfo{author}{Y.~Hu},
\newblock \bibinfo{title}{Experimental research on compressive mechanical properties of ice under low strain rates},
\newblock \bibinfo{journal}{Materials Today Communications} \bibinfo{volume}{24} (\bibinfo{year}{2020}) \bibinfo{pages}{101029}.
\bibitem[{Colgan et~al.(2016)Colgan, Rajaram, Abdalati, McCutchan, Mottram, Moussavi, and Grigsby}]{colgan2016glacier}
\bibinfo{author}{W.~Colgan}, \bibinfo{author}{H.~Rajaram}, \bibinfo{author}{W.~Abdalati}, \bibinfo{author}{C.~McCutchan}, \bibinfo{author}{R.~Mottram}, \bibinfo{author}{M.~S. Moussavi}, \bibinfo{author}{S.~Grigsby},
\newblock \bibinfo{title}{Glacier crevasses: Observations, models, and mass balance implications},
\newblock \bibinfo{journal}{Reviews of geophysics} \bibinfo{volume}{54} (\bibinfo{year}{2016}) \bibinfo{pages}{119--161}.
\bibitem[{Ren and Leslie(2014)}]{ren2014effects}
\bibinfo{author}{D.~Ren}, \bibinfo{author}{L.~M. Leslie},
\newblock \bibinfo{title}{Effects of waves on tabular ice-shelf calving},
\newblock \bibinfo{journal}{Earth Interactions} \bibinfo{volume}{18} (\bibinfo{year}{2014}) \bibinfo{pages}{1--28}.
\bibitem[{Hulbe et~al.(2016)Hulbe, Klinger, Masterson, Catania, Cruikshank, and Bugni}]{hulbe2016tidal}
\bibinfo{author}{C.~L. Hulbe}, \bibinfo{author}{M.~Klinger}, \bibinfo{author}{M.~Masterson}, \bibinfo{author}{G.~Catania}, \bibinfo{author}{K.~Cruikshank}, \bibinfo{author}{A.~Bugni},
\newblock \bibinfo{title}{Tidal bending and strand cracks at the kamb ice stream grounding line, west antarctica},
\newblock \bibinfo{journal}{Journal of Glaciology} \bibinfo{volume}{62} (\bibinfo{year}{2016}) \bibinfo{pages}{816--824}.
\bibitem[{Duddu and Waisman(2012)}]{duddu2012temperature}
\bibinfo{author}{R.~Duddu}, \bibinfo{author}{H.~Waisman},
\newblock \bibinfo{title}{A temperature-dependent creep damage model for polycrystalline ice},
\newblock \bibinfo{journal}{Mechanics of Materials} \bibinfo{volume}{46} (\bibinfo{year}{2012}) \bibinfo{pages}{23--41}.
\bibitem[{Hawkes and Mellor(1972)}]{hawkes1972deformation}
\bibinfo{author}{I.~Hawkes}, \bibinfo{author}{M.~Mellor},
\newblock \bibinfo{title}{Deformation and fracture of ice under uniaxial stress},
\newblock \bibinfo{journal}{Journal of glaciology} \bibinfo{volume}{11} (\bibinfo{year}{1972}) \bibinfo{pages}{103--131}.
\bibitem[{Murat and Lainey(1982)}]{murat1982some}
\bibinfo{author}{J.-R. Murat}, \bibinfo{author}{L.~Lainey},
\newblock \bibinfo{title}{Some experimental observations on the poisson's ratio of sea-ice},
\newblock \bibinfo{journal}{Cold Regions Science and Technology} \bibinfo{volume}{6} (\bibinfo{year}{1982}) \bibinfo{pages}{105--113}.
\bibitem[{Jimenez and Duddu(2018)}]{jimenez2018evaluation}
\bibinfo{author}{S.~Jimenez}, \bibinfo{author}{R.~Duddu},
\newblock \bibinfo{title}{On the evaluation of the stress intensity factor in calving models using linear elastic fracture mechanics},
\newblock \bibinfo{journal}{Journal of Glaciology} \bibinfo{volume}{64} (\bibinfo{year}{2018}) \bibinfo{pages}{759--770}.
\bibitem[{Liu and Miller(1979)}]{liu1979fracture}
\bibinfo{author}{H.~Liu}, \bibinfo{author}{K.~Miller},
\newblock \bibinfo{title}{Fracture toughness of fresh-water ice},
\newblock \bibinfo{journal}{Journal of glaciology} \bibinfo{volume}{22} (\bibinfo{year}{1979}) \bibinfo{pages}{135--143}.
\bibitem[{Petrovic(2003)}]{petrovic2003review}
\bibinfo{author}{J.~Petrovic},
\newblock \bibinfo{title}{Review mechanical properties of ice and snow},
\newblock \bibinfo{journal}{Journal of materials science} \bibinfo{volume}{38} (\bibinfo{year}{2003}) \bibinfo{pages}{1--6}.
\bibitem[{Zeng et~al.(2025)Zeng, Guilleminot, and Dolbow}]{dolbow2025uniform}
\bibinfo{author}{B.~Zeng}, \bibinfo{author}{J.~Guilleminot}, \bibinfo{author}{J.~E. Dolbow},
\newblock \bibinfo{title}{Examining crack nucleation under spatially uniform stress states with a complete phase-field model for fracture},
\newblock \bibinfo{journal}{Theoretical and Applied Fracture Mechanics}  (\bibinfo{year}{2025}) \bibinfo{pages}{105170}.
\bibitem[{Chockalingam et~al.(2026)Chockalingam, Tepole, and Kumar}]{chockalingam2025MCHB}
\bibinfo{author}{S.~Chockalingam}, \bibinfo{author}{A.~B. Tepole}, \bibinfo{author}{A.~Kumar},
\newblock \bibinfo{title}{The phase-field model of fracture incorporating mohr-coulomb, mogi-coulomb, and hoek-brown strength surfaces},
\newblock \bibinfo{journal}{Engineering Fracture Mechanics} \bibinfo{volume}{340} (\bibinfo{year}{2026}) \bibinfo{pages}{112108}.
\bibitem[{Nguyen et~al.(2025)Nguyen, Gupta, Duddu, and Annavarapu}]{duddu2025bodyforce}
\bibinfo{author}{D.~T. Nguyen}, \bibinfo{author}{A.~Gupta}, \bibinfo{author}{R.~Duddu}, \bibinfo{author}{C.~Annavarapu},
\newblock \bibinfo{title}{An adaptive mesh refinement algorithm for stress-based phase field fracture models for heterogeneous media: Application using fenics to ice-rock cliff failures},
\newblock \bibinfo{journal}{Finite Elements in Analysis and Design} \bibinfo{volume}{244} (\bibinfo{year}{2025}) \bibinfo{pages}{104311}.
\bibitem[{Lopez-Pamies and Kamarei(2025)}]{lopez2025Whenandwhere}
\bibinfo{author}{O.~Lopez-Pamies}, \bibinfo{author}{F.~Kamarei},
\newblock \bibinfo{title}{When and where do large cracks grow? {G}riffith energy competition constrained by material strength},
\newblock \bibinfo{journal}{Extreme Mechanics Letters} \bibinfo{volume}{81} (\bibinfo{year}{2025}) \bibinfo{pages}{102417}.
\bibitem[{Amor et~al.(2009)Amor, Marigo, and Maurini}]{AmorMarigoMaurini2009}
\bibinfo{author}{H.~Amor}, \bibinfo{author}{J.-J. Marigo}, \bibinfo{author}{C.~Maurini},
\newblock \bibinfo{title}{Regularized formulation of the variational brittle fracture with unilateral contact: Numerical experiments},
\newblock \bibinfo{journal}{Journal of the Mechanics and Physics of Solids} \bibinfo{volume}{57} (\bibinfo{year}{2009}) \bibinfo{pages}{1209--1229}.
\bibitem[{Kalthoff and Winkler(1988)}]{kalthoffwinkler1988}
\bibinfo{author}{J.~F. Kalthoff}, \bibinfo{author}{S.~Winkler},
\newblock \bibinfo{title}{Failure mode transition at high rates of shear loading},
\newblock \bibinfo{journal}{Impact loading and dynamic behaviour of materials} \bibinfo{volume}{1} (\bibinfo{year}{1988}) \bibinfo{pages}{185--195}.
\bibitem[{Ravi-Chandar and Knauss(1984{\natexlab{a}})}]{RaviChandar1984_SteadyStateCrack}
\bibinfo{author}{K.~Ravi-Chandar}, \bibinfo{author}{W.~Knauss},
\newblock \bibinfo{title}{An experimental investigation into dynamic fracture: Iii. on steady-state crack propagation and crack branching},
\newblock \bibinfo{journal}{International Journal of Fracture} \bibinfo{volume}{26} (\bibinfo{year}{1984}{\natexlab{a}}) \bibinfo{pages}{141--154}. \DOIprefix\doi{10.1007/BF01157550}.
\bibitem[{Ravi-Chandar and Knauss(1984{\natexlab{b}})}]{RaviChandar1984_CrackInitiationArrest}
\bibinfo{author}{K.~Ravi-Chandar}, \bibinfo{author}{W.~Knauss},
\newblock \bibinfo{title}{An experimental investigation into dynamic fracture: I. crack initiation and arrest},
\newblock \bibinfo{journal}{International Journal of Fracture} \bibinfo{volume}{25} (\bibinfo{year}{1984}{\natexlab{b}}) \bibinfo{pages}{247--262}. \DOIprefix\doi{10.1007/BF00963460}.
\bibitem[{Ravi-Chandar and Knauss(1984{\natexlab{c}})}]{RaviChandar1984_MicrostructuralAspects}
\bibinfo{author}{K.~Ravi-Chandar}, \bibinfo{author}{W.~Knauss},
\newblock \bibinfo{title}{An experimental investigation into dynamic fracture: Ii. microstructural aspects},
\newblock \bibinfo{journal}{International Journal of Fracture} \bibinfo{volume}{26} (\bibinfo{year}{1984}{\natexlab{c}}) \bibinfo{pages}{65--80}. \DOIprefix\doi{10.1007/BF01152313}.
\bibitem[{Ravi-Chandar(1998)}]{RaviChandar1998_DynamicFractureBrittle}
\bibinfo{author}{K.~Ravi-Chandar},
\newblock \bibinfo{title}{Dynamic fracture of nominally brittle materials},
\newblock \bibinfo{journal}{International Journal of Fracture} \bibinfo{volume}{90} (\bibinfo{year}{1998}) \bibinfo{pages}{83--102}. \DOIprefix\doi{10.1023/A:1007432017290}.
\bibitem[{Ravi-Chandar and Knauss(1984)}]{RaviChandar1984_ExperimentalDynamicFracture}
\bibinfo{author}{K.~Ravi-Chandar}, \bibinfo{author}{W.~Knauss},
\newblock \bibinfo{title}{An experimental investigation into dynamic fracture: Iv. on the interaction of stress waves with propagating cracks},
\newblock \bibinfo{journal}{International Journal of Fracture} \bibinfo{volume}{26} (\bibinfo{year}{1984}) \bibinfo{pages}{189--200}. \DOIprefix\doi{10.1007/BF01140627}.
\bibitem[{Ramulu and Kobayashi(1985)}]{Ramulu1985_MechanicsCrackCurving}
\bibinfo{author}{M.~Ramulu}, \bibinfo{author}{A.~Kobayashi},
\newblock \bibinfo{title}{Mechanics of crack curving and branching - a dynamic fracture analysis},
\newblock \bibinfo{journal}{International Journal of Fracture} \bibinfo{volume}{27} (\bibinfo{year}{1985}) \bibinfo{pages}{187--201}. \DOIprefix\doi{10.1007/BF00017967}.
\bibitem[{Belytschko et~al.(2003)Belytschko, Chen, Xu, and Zi}]{Belytschko2003_DynamicCrackHyperbolicity}
\bibinfo{author}{T.~Belytschko}, \bibinfo{author}{H.~Chen}, \bibinfo{author}{J.~Xu}, \bibinfo{author}{G.~Zi},
\newblock \bibinfo{title}{Dynamic crack propagation based on loss of hyperbolicity and a new discontinuous enrichment},
\newblock \bibinfo{journal}{International Journal for Numerical Methods in Engineering} \bibinfo{volume}{58} (\bibinfo{year}{2003}) \bibinfo{pages}{1873--1905}. \DOIprefix\doi{10.1002/nme.941}.
\bibitem[{Guo and Gao(2019)}]{Guo2019_KalthoffWinklerPeridynamics}
\bibinfo{author}{J.~Guo}, \bibinfo{author}{W.~Gao},
\newblock \bibinfo{title}{Study of the kalthoff–winkler experiment using an ordinary state-based peridynamic model under low velocity impact},
\newblock \bibinfo{journal}{Advances in Mechanical Engineering} \bibinfo{volume}{11} (\bibinfo{year}{2019}) \bibinfo{pages}{1--11}. \DOIprefix\doi{10.1177/1687814019852561}.
\bibitem[{Dahal and Kumar(2025)}]{dahal2025failure}
\bibinfo{author}{A.~Dahal}, \bibinfo{author}{A.~Kumar},
\newblock \bibinfo{title}{On failure mechanisms and load-parallel cracking in confined elastomeric layers},
\newblock \bibinfo{journal}{Extreme Mechanics Letters}  (\bibinfo{year}{2025}) \bibinfo{pages}{102406}.
\bibitem[{Gupta et~al.(2024)Gupta, Nguyen, and Duddu}]{gupta2024adaptive}
\bibinfo{author}{A.~Gupta}, \bibinfo{author}{D.~T. Nguyen}, \bibinfo{author}{R.~Duddu},
\newblock \bibinfo{title}{Damage mechanics challenge: Predictions from an adaptive finite element implementation of the stress-based phase-field fracture model},
\newblock \bibinfo{journal}{Engineering Fracture Mechanics} \bibinfo{volume}{306} (\bibinfo{year}{2024}) \bibinfo{pages}{110252}.
\bibitem[{Nguyen et~al.(2025)Nguyen, Gupta, Duddu, and Annavarapu}]{nguyen2025adaptive}
\bibinfo{author}{D.~T. Nguyen}, \bibinfo{author}{A.~Gupta}, \bibinfo{author}{R.~Duddu}, \bibinfo{author}{C.~Annavarapu},
\newblock \bibinfo{title}{An adaptive mesh refinement algorithm for stress-based phase field fracture models for heterogeneous media: Application using fenics to ice-rock cliff failures},
\newblock \bibinfo{journal}{Finite Elements in Analysis and Design} \bibinfo{volume}{244} (\bibinfo{year}{2025}) \bibinfo{pages}{104311}.
\bibitem[{Vella and Wettlaufer(2007)}]{vella2007finger}
\bibinfo{author}{D.~Vella}, \bibinfo{author}{J.~Wettlaufer},
\newblock \bibinfo{title}{Finger rafting: a generic instability of floating elastic sheets},
\newblock \bibinfo{journal}{Physical review letters} \bibinfo{volume}{98} (\bibinfo{year}{2007}) \bibinfo{pages}{088303}.

\end{thebibliography}

\end{document}